\documentclass{article}

\usepackage{amsmath}
\usepackage{amssymb}
\usepackage{authblk}
\usepackage{color}
\usepackage{graphicx}
\usepackage{hyperref}
\usepackage{lineno}
\usepackage{natbib}
\usepackage{setspace}
\usepackage{url}

\title{Detection of slow slip events using wavelet analysis of GNSS recordings}
\author[1]{Ariane Ducellier}
\author[2]{Kenneth C. Creager}
\author[2]{David A. Schmidt}
\affil[1]{Corresponding author. University of Washington, Department of Earth and Space Sciences, Box 351310, 4000 15th Avenue NE Seattle, WA 98195-1310, \href{mailto:ariane.ducellier.pro@gmail.com}{ariane.ducellier.pro@gmail.com}}
\affil[2]{University of Washington, Department of Earth and Space Sciences}
\date{}

\begin{document}

\maketitle

%
%

\doublespacing

\section*{Abstract}

In many places, tectonic tremor is observed in relation to slow slip and can be used as a proxy to study slow slip events of moderate magnitude where surface deformation is hidden in Global Navigation Satellite System (GNSS) noise. However, in subduction zones where no clear relationship between tremor and slow slip occurrence is observed, these methods cannot be applied, and we need other methods to be able to better detect and quantify slow slip. Wavelets methods such as the Discrete Wavelet Transform (DWT) and the Maximal Overlap Discrete Wavelet Transform (MODWT) are mathematical tools for analyzing time series simultaneously in the time and the frequency domain by observing how weighted differences of a time series vary from one period to the next. In this paper, we use wavelet methods to analyze GNSS time series and seismic recordings of slow slip events in Cascadia. We use detrended GNSS data, apply the MODWT transform and stack the wavelet details over several nearby GNSS stations. As an independent check on the timing of slow slip events, we also compute the cumulative number of tremor in the vicinity of the GNSS stations, detrend this signal, and apply the MODWT transform. In both time series, we can then see simultaneous waveforms whose timing corresponds to the timing of slow slip events. We assume that there is a slow slip event whenever there is a positive peak followed by a negative peak in the wavelet signal. We verify that there is a good agreement between slow slip events detected with only GNSS data, and slow slip events detected with only tremor data for northern Cascadia. The wavelet-based detection method effectively detects events of magnitude higher than 6 as determined by independent event catalogs (e.g. ~\citep{MIC_2019}). As a demonstration of using the wavelet analysis in a region without significant tremor, we also analyze GNSS data from New Zealand and detect slow slip events that are spatially and temporally close to those detected previously by other studies.

\section{Introduction}

Slow slip events are new phenomena discovered in the last two decades in many subduction zones thanks to recordings of the displacement of Earth's surface by dense Global Navigation Satellite System (GNSS) networks ~\citep{VER_2010,SCH_2010,JIA_2012,WAL_2012}. As with ordinary earthquakes, slow slip events represent slip on a fault, for instance the plate boundary between a tectonic plate subducting under another tectonic plate. However, they take a much longer time (several days to several years) to happen relative to ordinary earthquakes. They have a relatively short recurrence time (months to years) compared to the recurrence time of regular earthquakes (up to several hundreds of years), allowing scientists to observe and study many complete event cycles, which is typically not possible to explore with traditional earthquake catalogs ~\citep{BER_2011}. A slow slip event on the plate boundary is inferred to happen when there is a reversal of the direction of motion at GNSS stations, compared to the secular interseismic motion. Slow slip events have been observed in many places ~\citep{BER_2011,AUD_2016}, such as Cascadia ~\citep{BAR_2020}, Nankai ~\citep{NIS_2013}, Alaska ~\citep{LI_2016}, Costa Rica ~\citep{JIA_2012}, Mexico ~\citep{RAD_2012}, and New Zealand ~\citep{WAL_2020}. \\

In many places, tectonic tremor is also observed in relation to slow slip, but the spatial agreement between tremor and slow slip may vary along the strike of the plate boundary ~\citep{HAL_2018}. Tremor is a long (several seconds to many minutes), low amplitude seismic signal, with emergent onsets, and an absence of clear impulsive phases. Tectonic tremor have been explained as a swarm of small, low-frequency earthquakes (LFEs) ~\citep{SHE_2007_nature}, which are small magnitude earthquakes (M $\sim$ 1) whose frequency content (1-10 Hz) is lower than for ordinary earthquakes (up to 20 Hz). In subduction zones such as Nankai and Cascadia, tectonic tremor observations agree spatially and temporally with slow slip observations ~\citep{ROG_2003,OBA_2004}. Due to this agreement, these paired phenomena have been called Episodic Tremor and Slip (ETS). However, this is not always the case. For instance, in northern New Zealand, tremor is more challenging to detect, and seems to be located downdip of the slow slip on the plate boundary ~\citep{TOD_2016}. In Alaska, the tremor zone only partially overlaps the long-term slow slip zone and there does not appear to be any temporal agreement between tremor and slow slip occurrence ~\citep{WEC_2016}. \\

In Cascadia, there are robust signals in both slow slip and tremor ~\citep{HAW_2013}. This is also the case in Nankai ~\citep{HIR_2008}, where tiltmeters are used instead of GNSS. It is thus possible to use tremor as a proxy to observe slow slip events that are not directly observed in the GNSS data. For instance, ~\citet{AGU_2009} studied 23 ETS events in Cascadia with more than 50 hours of tectonic tremor. For all these events, they computed both the GPS-estimated moment release and the cumulative number of hours of tectonic tremor recorded. They observed a linear relationship between moment release and number of hours of tremor for slow slip events of moment magnitude 6.3 to 6.8. Based on this linear relationship, it is possible to infer the existence of smaller slow slip events of magnitude 5-6 occurring simultaneously with smaller tremor bursts of duration 1 to 50 hours occurring in between the big ETS events, and for which there is no detectable signal in the GPS data. \\

~\citet{FRA_2016} divided GPS time series observations from Cascadia and Guerrero, Mexico, into two groups: the first group contains days with abundant tremor and LFEs, the second group contains days when the number of tremor or LFEs is lower than a threshold. He then stacked separately the two groups of daily observations and observed a cumulative displacement in the direction corresponding to the loading period when few tremor or LFEs are observed and the surface deformation corresponds to the secular plate motion. He also observed a cumulative displacement in the opposite direction corresponding to the release period when tremor and LFEs are observed. He was thus able to observe a reverse displacement corresponding to smaller slow slip events not directly observable in the GPS data for individual events. \\

However, these methods cannot be applied to detect slow slip events in places where tremor and slow slip occurrence are not well spatially and temporally correlated, tremor is not abundant, or the seismic network is not robust enough. We thus need other methods to be able to better detect and quantify slow slip. \\

Wavelet methods such as the Discrete Wavelet Transform (DWT) are mathematical tools for analyzing time series simultaneously in the time and the frequency domain by observing how weighted differences of a time series vary from one period to the next. Wavelet methods have been widely used for geophysical applications (e.g. ~\citet{KUM_1997}). However, few studies have used wavelet methods to analyze recordings of slow slip, and their scope was limited to the detection of the bigger (magnitude 6-7) short-term (a few weeks) events ~\citep{SZE_2008,OHT_2010,WEI_2012,ALB_2019}. \\

~\citet{SZE_2008} determined the timing and the amplitude of 34 slow slip events throughout the Cascadia subduction zone between 1997 and 2005 using wavelets. They modeled the GPS time series by the sum of a linear trend, annual and biannual sinusoids representing seasonal effects, Heaviside step functions corresponding to earthquakes and hardware upgrades, and a residual signal. They then applied a Gaussian wavelet transform to the residual time series to get the exact timing of slow slip at each GPS station. The idea is that the wavelet transform allows us to analyze the signal both in the time and the frequency domains. A sharp change in the signal will be localized and seen at all time scales of the wavelet decomposition, contrary to what happens with the periodic sinusoids of the Fourier transform. \\

Instead of using wavelets in the time domain, ~\citet{OHT_2010} used 2D wavelet functions in the spatial domain to detect slow slip events. They designed the Network Stain Filter (NSF) to detect transient deformation signals from large-scale geodetic arrays. They modeled the position of the GPS station by the sum of the secular velocity, a spatially coherent field, site-specific noise, reference frame errors, and observation errors. The spatial displacement field is modeled by the sum of basis wavelets with time-varying weights.  Their method has been successfully used to detect a transient event in the Boso peninsula, Japan, and a slow slip event in the Alaska subduction zone ~\citep{WEI_2012}. \\

Finally, ~\citet{ALB_2019} used hourly water level records from four tide gauges in the Juan de Fuca Straight and the Puget Sound to determine relative vertical displacements associated with slow slip events between 1996 and 2011. Their main idea is that the tidal level measured at a given gauge is the sum of a noise component at multiple timescales (tides, ocean and atmospheric noise) and an uplift signal due to the slow slip events. The noise component is assumed to be coherent between all tidal gauges, while the tectonic uplift signal is different provided that the gauges are far enough from each other. By stacking the tidal records after removing tides, the uplift signals cancel each other while the noise signal is amplified. By stacking the components at different time scales of the DWT decomposition, instead of stacking the raw tidal record, each of the components of the noise at different time scales is retrieved and can then be removed from the raw records to obtain the uplift signal. Due to the relative location of the tidal gauges at Port Angeles and Port Townsend compared to the slow slip region on the plate boundary, a slow slip event should result in uplift in Port Angeles (western part) and in subsidence in Port Townsend (eastern part). Indeed, the authors were able to clearly see a difference in the sign of the uplift at these two tidal gauges. \\

In our study, we use a similar approach to previous studies with a different reasoning. We only stack signals at nearby GPS stations, assuming that the east-west displacement due to the slow slip events will then be the same at each of the GPS stations considered. We suppose that some of the noise component is different at each GPS station and will be eliminated by the stacking. Finally, we assume that the noise and the longitudinal displacement due to the slow slip events and the secular plate motion have different time scales, so that the wavelet decomposition will act as a bandpass filter to retrieve the displacement signal and highlight the slow slip events. We use wavelet methods to analyze GPS and tremor recordings of slow slip events in Cascadia. Our objective is to verify that there is a good agreement between slow slip events detected with only GNSS data, and slow slip events detected with only tremor data. We thus want to demonstrate that the wavelet-based detection method can be applied to detect slow slip events that may currently be obscured using standard methods. Finally, we apply the method to GNSS data in New Zealand and successfully detect several slow slip events without needing to rely on the tremor data. \\

\section{Data}

We first focused our study on northwest Washington State. For the GNSS data, we used the GPS time series provided by the Pacific Northwest Geodetic Array, Central Washington University. These are network solutions in ITRF2014 with phase ambiguities resolved with wide-lane phase-biases. Orbits and satellite clocks provided by the Jet Propulsion Laboratory/NASA. North, East, and Vertical directions are available. However, as the direction of the secular plate motion is close to the East direction, we only used the East direction of the GPS time series for the data analysis, as it has the best signal-to-noise ratio. The wavelet method works best with data with zero mean, and no sharp discontinuities; so we use the cleaned dataset, that is GPS times series with linear trends, steps due to earthquakes or hardware upgrades, and annual and semi-annual sinusoids signals simultaneously estimated and removed following ~\citet{SZE_2004}. For the tremor data, we used the tremor catalog from the Pacific Northwest Seismic Network (PNSN) ~\citep{WEC_2010}. \\

For the application to slow slip events in New Zealand, we used the GPS time series provided by the Geological hazard information for New Zealand (GeoNet). The coordinates have been extracted by GeoNet during the GLOBK run from the combined daily solution files, and converted to (east, north, up) displacement in millimeters with respect to an a priori position and epoch in the ITRF2008 realization. The time series provided by GeoNet have no adjustments made to them, so they may, for example, contain offsets due to earthquakes, offsets due to equipment changes at individual sites, and seasonal (annual and semi-annual) signals due to various causes. Here again, the direction of the secular interseismic plate motion is close to the West direction, so we only used the East-West component of the GPS time series for the data analysis. We detrended the data before applying the wavelet transform by carrying a linear regression of the whole time series and removing the straight line obtained from the regression.

\section{Method} 

\subsection{The Maximal Overlap Discrete Wavelet Transform}

The Discrete Wavelet Transform (DWT) is an orthonormal transform that transforms a time series $X_t \left( t = 0, \cdots , N - 1 \right)$ into a vector of wavelet coefficients $W_i \left( i = 0 , \cdots , N - 1 \right)$. If we denote $J$ the level of the wavelet decomposition, and the number of observations is equal to $N = n * 2^J$, where $n$ is some integer greater than or equal to 1, the vector of wavelet coefficients can be decomposed into $J$ wavelet vectors $W_j$ of lengths $\frac{N}{2}$, $\frac{N}{4}$, ... , $\frac{N}{2^J}$, and one scaling vector $V_J$ of length $\frac{N}{2^J}$. Each wavelet vector $W_j$ is associated with changes on time scale $\tau_j = dt 2^{j - 1}$, where $dt$ is the time step of the time series, and corresponds to the filtering of the original time series with a filter with nominal frequency interval $\lbrack \frac{1}{dt 2^{j + 1}} ; \frac{1}{dt 2^j} \rbrack$. The scaling vector $V_J$ is associated with averages in time scale $\lambda_J = dt 2^J$, and corresponds to the filtering of the original time series with a filter with nominal frequency interval $\lbrack 0 ; \frac{1}{dt 2^{j + 1}} \rbrack$. Wavelet vectors can be further decomposed into details and smooths, which are more easily interpretable. We define for $j = 1 , \cdots , J$ the $j$th wavelet detail $D_j$, which is a vector of length $N$, and is associated to time scale $\tau_j = dt 2^{j - 1}$. Similarly, we can define for $j = 1 , \cdots , J$ the $j$th wavelet smooth $S_j$, which is a vector of length $N$, and is associated to scales $\tau_{j + 1} = dt 2^{j + 1}$ and higher. The basic idea is to reapply to $W_j$ the wavelet filter that was used to construct $W_j$ from the initial time series $X$. Together, the details and the smooths define the multiresolution analysis (MRA) of $X$:

\begin{linenomath*}
\begin{equation}
X = \sum_{j = 1}^{J} D_j + S_J
\end{equation}
\end{linenomath*}

The DWT presents several disadvantages. First, the length of the time series must be a multiple of $2^J$ where $J$ is the level of the DWT decomposition. Second, the time step of the wavelet vector $W_j$ is $dt 2^j$, which may not correspond to the time when some interesting phenomenon is visible on the original time series. Third, when we circularly shift the time series, the corresponding wavelet coefficients, details and smooths are not a circularly shifted version of the wavelet coefficients, details and smooths of the original time series. Thus, the values of the wavelet coefficients, details and smooths are strongly dependent on the time when we start experimentally gathering the data. Finally, when we filter the time series to obtain the details $D_j$ and smooths $S_j$, we introduce a phase shift, which makes it difficult to line up meaningfully the features of the MRA with the original time series. \\

To overcome the disadvantages described above, we use instead the Maximal Overlap Discrete Wavelet Transform (MODWT). The MODWT transforms the time series $X_t \left( t = 0, ... , N - 1 \right)$ into J wavelet vectors $\widetilde{W}_j \left( j = 1 ,  \cdots , J \right)$ of length $N$ and a scaling vector $\widetilde{V}_J$ of length $N$. As is the case for the DWT, each wavelet vector $\widetilde{W}_j$ is associated with changes on scale $\tau_j = dt 2^{j - 1}$, and corresponds to the filtering of the original time series with a filter with nominal frequency interval $\lbrack \frac{1}{dt 2^{j + 1}} ; \frac{1}{dt 2^j} \rbrack$. The scaling vector $\widetilde{V}_J$ is associated with averages in scale $\lambda_J = dt 2^J$, and corresponds to the filtering of the original time series with a filter with nominal frequency interval $\lbrack 0 ; \frac{1}{dt 2^{j + 1}} \rbrack$. As is the case for the DWT, we can write the MRA:

\begin{linenomath*}
\begin{equation}
X = \sum_{j = 1}^{J} \widetilde{D}_j + \widetilde{S}_J
\end{equation}
\end{linenomath*}

The MODWT of a time series can be defined for any length $N$. The time step of the wavelet vectors $\widetilde{W}_j$ and the scaling vector $\widetilde{V}_J$ is equal to the time step of the original time series. When we circularly shift the time series, the corresponding wavelet vectors, scaling vector, details and smooths are shifted by the same amount. The details and smooths are associated with a zero phase filter, making it easy to line up meaningfully the features of the MRA with the original time series. The wavelet methods for time series analysis are explained in a more detailed way in ~\citep{PER_2000}). \\

The boundary conditions at the two edges of the time series will affect the wavelet coefficients. For the MODWT, if we denote $L$ the length of the base wavelet filter used for the wavelet decomposition (in our study, we used a Least Asymmetric wavelet filter of length $L = 8$, see ~\citep{PER_2000}, section 4.8, page 107), the length of the wavelet filter at level $j$ used to compute the wavelet detail $D_j$ is:

\begin{equation*}
L_j = \left( 2^j - 1 \right) \left( L - 1 \right) + 1
\end{equation*}

The wavelet coefficients of the detail al level $j$ affected by the boundary conditions at the edges would then be the coefficients with indices $t = 0 , \cdots , L_j - 2$ or $t = N - L_j + 1 , \cdots, N - 1$ (see ~\citep{PER_2000}, section 5.11, page 199). We get $L_j = 442$ for $j = 6$, $L_j = 890$ for $j = 7$ and $L_j = 1786$ for $j = 8$. In practice, the part of the wavelet details affected by the boundary conditions is much shorter than that. We compared the wavelet details computed when using only the data between 2008 and 2012 and the wavelet details computed when using the entire time series from 2000 to 2021 (Figure S1 in the Supplementary Material). Even at level 8 only about 6 months of data on each side are effected by the boundary conditions.

\subsection{Application to synthetic data}

To illustrate the wavelet transform method, we first apply the MODWT to synthetic data. As slow slip events occur in Cascadia on a regular basis, every twelve to eighteen months, we create a synthetic signal of period $T = 500$ days. To reproduce the ground displacement observed on the longitudinal component of GPS stations in Cascadia, we divide each period into two parts: In the first part of duration $T - N$, the displacement is linearly increasing and corresponds to the inter seismic plate motion in the eastern direction; in the second part of duration $N$, the displacement is linearly decreasing and corresponds to a slow slip event on a reverse fault at depth triggering a ground displacement in the western direction. To see the effect of the duration of the slow slip event, we use different values for $N = 5, 10, 20, 40$ days. The amplitude of the set is normalized to 1. Figure 1 shows the synthetics, the details $D_j$  of the wavelet decomposition for levels 1 to 10, and the smooth $S_{10}$ for the four durations of a slow slip event. \\

The ramp-like signal is transformed through the wavelet filtering into a waveform with first a positive peak and then a negative peak. The shape of the waveform is the same for every level of the wavelet decomposition, but the width of the waveform increases with the scale level. For the 8th level of the wavelet decomposition, the width of the waveform is nearly as large as the time between two events. At larger scales, the waveforms start to merge two contiguous events together, and make the wavelet decomposition less interpretable. For an event of duration 5 days, the wavelet details at levels higher than 3 have a larger amplitude than the wavelet details at lower scales. For an event of duration 10 days, the wavelet details at levels higher than 4 have a larger amplitude than the wavelet details at lower scales. For an event of duration 20 days, the wavelet details at levels higher than 5 have a larger amplitude than the wavelet details at lower scales. For an event of duration 40 days, the wavelet details at levels higher than 6 have a larger amplitude than the wavelet details at lower scales. Thus, the scale levels at which an event is being seen in the wavelet details give us an indication about the duration (and the magnitude) of the slow slip event. The big slow slip events of magnitude 6-7 typically trigger a signal that lasts about one week at an individual GPS station, and the whole event lasts several weeks. We expect them to start being visible at the level 5 of the wavelet decomposition, but to not be noticeable at lower time scales. \\

\subsection{MODWT of GPS and tremor data}

The DWT and MODWT methods must be used on a continuous time series, without gaps in the recordings. To deal with the gaps in the GNSS recordings, we simply replace the missing values by interpolation. The value for the first day for which data are missing is equal to the mean of the five days before the gap. The value for the last day for which data are missing is equal to the mean of the five days after the gap. The remaining missing values are computed by doing a linear interpolation of the first and the last values and adding a Gaussian noise component with mean zero and standard deviation equal to the standard deviation of the whole time series. We verify how the wavelet details may be affected by looking at a GPS time series without missing values and compared the wavelet details with and without removing some data points. Station PGC5 recorded continuous 1390 days between 2009 and 2013 without any missing values. We first computed the wavelet details without missing values. Then, we removed ten neighboring values, replaced them using the method described above (linear interpolation plus Gaussian noise), and computed the wavelet details with the replaced values. Figure S2 in the Supplementary Material shows a comparison of the two wavelet details for two different locations of the missing values. We can see that there are visible differences in the time series itself, and in the details at the smallest levels of the wavelet decomposition. However, the differences between the wavelet details with and without missing values get smaller and smaller with increasing levels of details, and are barely visible for the levels that are most relevant (levels 6 and above). We thus conclude that we can easily replace the missing values in the GNSS time series without introducing false detections of slow slip events. \\

We then applied the wavelet filtering to real GPS data. Figure 2 shows the longitudinal displacement for GPS station PGC5, located in southern Vancouver Island, the details of the wavelet decomposition for levels 1 to 8, and the smooth. In the data, we can see a sharp drop in displacement whenever there is a documented slow slip event. For levels 5 to 8, which correspond to time scales 16, 32, 64 and 128 days, we can see in the details a positive peak followed by a negative peak whenever there is a drop in displacement in the data. We thus verify that the wavelet method can detect steps in the time series associated with slow slip events. \\

To increase the signal-to-noise ratio and better detect slow slip events, we stack the signal from several neighboring GPS stations. We choose to focus on GPS stations located close enough to the tremor zone to get a sufficiently high amplitude of the slow slip signal. We choose 16 points along the 40 km depth contour of the plate boundary (model from ~\citet{PRE_2003}) with spacing equal 0.1 degree in latitude (red triangles on Figure 3). Then we took all the GPS stations located in a 50 km radius for a given point, compute the wavelet details for the longitudinal displacement of each station, and stack each detail over the GPS stations. We thus have a stacked detail for each level 1 to 10 of the wavelet decomposition. \\

To assess the success of the wavelet decomposition for detecting slow slip events in GPS time series, we validate the approach by comparing to an independent proxy for slow slip events. We took all the tremor epicenters located within a 50 km radius centered on one of the 16 locations marked by red triangles on Figure 3. Then we computed the cumulative number of tremor within this circle. Finally, we removed a linear trend from the cumulative tremor count, and applied the wavelet transform. Because of the preprocessing applied to the tremor data before that wavelet transform, the measurement unit associated with the corresponding wavelet details is the fraction of tremor in a day divided by the total number of days. The average value is 1 divided by the total number of days. Figure 4 shows an example of the wavelet decomposition for the third northernmost location on Figure 3 (which is closest to GPS station PGC5). Contrary to what happens for the GPS data, we see a sharp increase in the time series whenever there is a tremor episode, which translates into a negative peak followed by a positive peak in the wavelet details.

\section{Application to data from Cascadia}

We stacked the 8th level detail of the wavelet decomposition of the displacement over all the GPS stations located in a 50 km radius of a given point, for the 16 locations indicated in Figure 3. The result is shown in the top panel of Figure 5, where each line represents one of the locations along strike. To better highlight the peaks in the wavelet details, we highlighted in red the time intervals where the amplitude of the stacked detail is higher than a threshold, and in blue the  time intervals where the amplitude of the stacked detail is lower than minus the threshold. To compare the GPS signal with the tremor signal, we plotted the 8th level detail of the wavelet decomposition of the tremor count on the bottom panel of Figure 5. We multiplied by -1 the cumulative tremor count for the wavelet decomposition in order to be able to match positive peaks with positive peaks and negative peaks with negative peaks. In the tremor catalog from the PNSN, there are 17 tremor events with more than 150 hours of tremor recorded. The events are summarized in Table 1. The time of the event is the start date plus half the duration of the event. \\

Although the latitudinal extension of the events is not always the same for the GPS data and for the tremor data, we identify the same 13 events in both 8th wavelet decompositions for the 8th level: January 2007, May 2008, May 2009, August 2010, August 2011, September 2012, September 2013, August-November 2014, January 2016, March 2017, June 2018, March-November 2019, and October 2020-January 2021. Although there are two events in the tremor catalog in August 2014 and November 2014, these two events are not distinguishable in the 8th level details and look more like a single event slowly propagating from South to North. The same phenomenon is observed in 2019 when two tremor events in March and November 2019 are merged into a single event propagating slowly from South to North. In 2020-2021, the wavelet decomposition of the tremor shows one event in the south in October-November 2020 and one event in the North in January 2021, but in the wavelet decomposition of the GPS data, these three events look like a single event propagating slowly from South to North. \\

A similar comparison is shown for the wavelet decomposition of the GPS data and the wavelet decomposition of the tremor count data for the 7th level and the 6th level respectively (Figures 6 and 7). The events are harder to see in the 7th level than in the 8th level, both for the GPS data and the tremor count data. The wavelet decomposition is more noisy for the GPS data between 2010 and 2012, but it does not seem that there are more slow slip events visible in the 7th level. \\

For the 6th level detail, we see an additional event in the South in Fall 2009 that is present both in the GPS and the tremor data. It may correspond to the northern extent of a big ETS event occurring in Fall 2009 south of the study area (event 19 in the ~\citet{MIC_2019} catalog). There are three small signals in the GPS data in Winter 2012, Fall 2017, and Winter 2020 that are not present in the tremor data, and may be false detections. To summarize, we assume that robust detections are events present in both GPS and tremor time series, and false detections are events present in the GPS but not in the tremor time series. Then, all the 13 events present on the 8th level detail of the wavelet decomposition are robust detections and 14 of the 17 events present on the 6th level detail of the wavelet decomposition are robust detections. \\

To better evaluate the number of robust and false detections, we convert the wavelet details into trinary time series. If the absolute value of the wavelet detail is higher that a threshold, we replace the value by 1 (for positive values) or -1 (for negative values), otherwise we replace the value by 0. We do this on both the wavelet details of the GPS data and of the tremor data. Then we decide that if both the GPS and the tremor time series take the value 1 (or both take the value -1), we have a robust detection (true positive, TP). If the GPS and the tremor time series have opposite signs, or if the absolute value of the GPS time series is 1 but the value of the tremor time series is 0, we have a false detection (false positive, FP). If both time series take the value 0, we do not have detection (true negative, TN). If the GPS time series take the value 0, but the absolute value of the tremor time series is 1, we miss a detection (false negative, FN). We then define the sensitivity (true positive rate) and the specificity (equal to 1 minus the false positive rate) as:

\begin{equation}
\begin{aligned}
\text{sensitivity} &= \frac{TP}{TP + FN} \\
\text{specificity} &= \frac{TN} {TN + FP}
\end{aligned}
\end{equation}

We can then evaluate the quality of the detections obtained with our method by plotting a receiver operating characteristic curve (ROC curve). The ROC curve is widely use for binary classification problems in statistics and machine learning. We calculate an ROC value by varying the values of the threshold (here the two thresholds used to convert the GPS and the tremor time series into trinary time series), computing the corresponding values of the true positive rate and the false positive rate (equal to 1 minus the specificity), and plotting the true positive rate as a function of the false positive rate. If the classification was made randomly, all the points would fall on the first diagonal. If the classifier was perfect, the corresponding point would fall on the top left corner of the graph with true positive rate equal to 1 and false positive rate equal to 0. The bigger the area under the curve, the better the classification method is. \\

As the slow slip events are better seen on levels 6, 7 and 8 of the wavelet decomposition, we first add the wavelet details corresponding to levels 6 to 8, and transform the resulting time series into a trinary time series. We apply this transform to both the GPS and the tremor time series with varying thresholds. We then plot the ROC curve on Figure 8, each dot representing a different threshold. The corresponding sums of the wavelet details for the GPS data and the tremor data are shown on Figure 9. We can see that there is a trade-off between sensitivity and specificity as we vary the threshold. If we decrease the false positive rate, we also decrease the number of true events detected. If we increase the number of true events detected, we also increase the false positive rate. If we increase the threshold for the tremor, the curve goes farther away from the first diagonal, that is we get better classification results. If we increase the threshold for the GPS, the false positive rate and the the number of events detected decrease. In Figure 9, we have chosen thresholds for the GPS time series and the tremor time series such that the specificity is higher than 0.75 (that is the false positive rate is lower than 0.25), and the sensitivity is the highest possible, that is we have chosen the thresholds corresponding to the dot that is farthest from the diagonal, which is random. \\

In addition to the magnitude 6 events discussed above,  ~\citet{MIC_2019} have also identified several magnitude 5 events using a variational Bayesian Independent Component Analysis (vbICA) decomposition of the signal. As we expect smaller magnitude events to be more visible at smaller time scales of the wavelet decomposition (level 5), we verify for all these events whether a signal can be seen at the same time as the time given in their catalog. Most of these magnitude 5 events are also sub-events of bigger magnitude 6 events. Table 2 summarizes for each event its timing, its number and its magnitude as indicated in the catalog from ~\citet{MIC_2019}, and whether it is part of a bigger magnitude 6 event. Figure 10 shows the 5th level detail wavelet decomposition of the GPS data. Red lines show the timing of the big slow slip events from Table 1, and blue lines show the timing of the small slow slip events from Table 2. \\

All 14 events that are sub-events of a bigger event are visible at level 5. However, this may be because the bigger events are clearly seen at levels 6 to 8, and also at smaller time scales. The one small event that is not part of a bigger event (Winter 2009) is visible at level 5 of the wavelet decomposition. However, some other events that are not in the catalog of ~\citet{MIC_2019}'s catalog are also visible in late 2007, early 2010, early 2012, and early 2020. Therefore, it is difficult to differentiate between a robust detection and a false detection, and to conclude whether the method can indeed detect events of magnitude 5. \\

In Figure 9, we see four smaller events that are not in the catalog of ~\citet{MIC_2019}: at about 2007.5, there is a negative peak followed by a positive peak (that is an event in the opposite direction of what would be expected from slow slip), at about 2010.2, 2012.2 and 2020.2, there are positive peaks followed by negative peaks for all the sixteen locations studied in this paper. These events are highlighted in Figure S4 in the Supplementary Information. Looking back at the original GPS data, there is a small increase in the displacement in the eastern direction that lasts about one or two months at about 2007.5. However, the direction of the displacement does not correspond to a slow slip event, and another cause should be found to explain this signal. There is a decrease in displacement that lasts several months at about 2010.2. This transient may correspond to a long duration slow slip event. There is a small decrease in displacement at about 2012.2. Its amplitude is small but the duration and direction correspond to a slow slip event, so this transient could be a very small slow slip event. Finally, there is also a small decrease in displacement at about 2020.2 that is difficult to interpret. \\

Due to the short distances between the GPS stations and the locations of the red triangles on the map from Figure 3, the same station could be used multiple times for the stacking at different locations. When considering two different locations, the stacking is thus made over an overlapping number of stations. Table 3 summarizes the number of stations and the number of overlapping stations for each location on Figure 3. We hypothesize that the small displacement in the eastern direction seen at about 2007.5 could be due to a misbehaving station common to several locations. However, several GPS stations indeed show an increase in the displacement in the eastern direction at about 2007.5. There are many missing data around that time, so it is difficult to conclude. \\

Another possibility is that common mode signals could stack constructively across GNSS stations and produce peaks in the wavelet details that are actually due to non-tectonic signals. We computed common mode signals for different latitude bins (each bin has width equal to half-a-degree of latitude) following the same method as \citet{NUY_2021}. We first stacked all the time series for the stations in each latitude bin that are located more than 100 km east of the 40 km depth contour of the plate boundary. We assume that these stations are not sensitive to the deformation on the plate interface. We then apply a yearly moving average to each common mode signal in order to remove any leftover noise. The common mode signal was then removed from the GNSS time series depending on each site’s latitude. Figure S3 in the Supplementary Information shows the corresponding sum of the stacks of the 6th, 7th and 8th wavelet details obtained from the resulting time series. The common modes seem to have little impact on the results and do not explain the additional four small events that we noted in Figure 9. \\

In order to convert our filtered eastward displacement time series into a slow slip event catalog we note that red bars represent displacements exceeding a threshold of 0.8 mm (east), and blue marks displacements less than minus -0.8 mm (west).  During times with no slow slip GPS stations on the overriding plate are pushed slowly eastward by the locked subducting plate.  Slow slip events represent GPS motion towards the west.  Thus, we infer that slow slip events happen when red bars are immediately followed by blue bars in the wavelet details.  We have identified everywhere that this has happened and mark it with a green line in Figure 11 and as a row in Table 4.  We find 17 possible SSEs by this method using filtered GPS data only. For each of these 17 events we determine the time difference between the mid time of the GPS catalog and the nearest time from the tremor catalog (Table 1). These time differences are in column 6 (Table 4).  Every event in the GPS catalog has a match in the tremor catalog except for the tremor event at 2010.15. There is also only one event in the tremor catalog that is not in the GPS catalog. It occurs at 2014.65 with a duration of 15 days and 190 hours of tremor. It occurs 0.25 years after the nearest GPS event. There are also two marginal events in the tremor catalog with time differences of 0.13 and 0.10 years, but those are also among the smaller events with 162 and 193 hours of tremor.

\section{Application to data from New Zealand}

We now apply our wavelet-based method to detect slow slip events in New Zealand, a location where the spatial and temporal agreement between tremor and slow slip is not as good as in other subduction zones. The tectonics of the North Island of New Zealand are dominated by the westward subduction of the Pacific Plate under the Australian Plate at the Hikurangi Trench. Two types of slow slip events have been observed at the Hikurangi margin. Shallow (10-15 km depth), shorter (1-3 weeks), and usually smaller (Mw 6.3-6.8) slow slip events have been observed every 18-24 months in the northern part of the margin. Deeper (35-60 km depth), longer (12-18 months), and larger (Mw 7.0) slow slip events have been observed every 5 years in the southern part of the margin ~\citep{WAL_2010,TOD_2016}. The detection of tremor has been elusive in northern Hikurangi. ~\citet{DEL_2009} observed an increase in the rate of microseismicity downdip of the 2004 Gisborne slow slip event. More recently, however, ~\citep{KIM_2011} detected a low level of tremor activity that increased during the 2010 Gisborne slow slip event. As was the case for the microearthquakes, the source of the tremor was located downdip of the slow slip patch determined from GNSS data. ~\citep{IDE_2012} detected tremor downdip of the location of two deep slow slip events observed by ~\citet{WAL_2013} in 2006 and 2008. However, contrary to ETS events in Cascadia and Nankai, the tremor activity did not seem to increase during the slow slip events. ~\citet{TOD_2016}  detected tremor associated with most of the shallow slow slip events between 2010 and 2015, and located downdip of the geodetically inferred slip area. They also detected deeper tremor between 20 and 50 km depth with unclear origin. They hypothesized that these tremor may be related to undetected deep long-term slow slip events. \\

To evaluate whether the wavelet analysis is effective in a region without robust tremor, we take all the New Zealand GPS stations located in a 50 km radius of a given location, for the 18 locations indicated in Figure 12, and we stack the 6th level details, the 7th level details or the 8th level details over all the GPS stations. We then sum together the 6th, 7th and 8th levels stacked wavelet details (Figure 13, top panel). We highlight positive and negative peaks with red and blue colors as was done in Figure 9. We cannot use the tremor data to decide what is the appropriate threshold above which we consider that there is a slow slip event. Slow slip events in New Zealand result in surface displacements that are similar in amplitude to twice as large as those observed in Cascadia. Therefore, the amplitudes of the peaks in the wavelet details should be similar in New Zealand and in Cascadia and we choose identical thresholds for both regions. As a slow slip event in northern New Zealand results in a displacement in the east direction at the earth’s surface, the slow slip events are indicated by a negative peak followed by a positive peak in the stacked wavelet details. We compare the results of the timings and locations of the slow slip events to those events detected by ~\citet{TOD_2016}. As they only used data from five GPS stations (PUKE, ANAU, GISB, MAHI and CKID), we indicate by a vertical orange bar on the bottom panel of Figure 13 each time a slow slip event was detected for these stations. The orange bars are centered on the latitudes of the GPS stations. If a slow slip event was detected by more than one station, all the corresponding orange bars are linked together to show the spatial extent of the slow slip. ~\citet{TOD_2016} indicated by a question mark (on their Figure 2 and their Table 1) additional possible events, and those are indicated by a dotted orange bar on Figure 13. To compare with the slow slip events detected with the wavelet method, we also mark by a green bar every time a negative peak lower than the threshold is followed by a positive peak higher than the threshold. Table 5 summarizes the slow slip events detected with the wavelet method for 2010-2016. \\

We observe that there is a good agreement between the events detected with the wavelet method and the events previously detected by ~\citet{TOD_2016}. We clearly see an event propagating from south to north in January-February (event 2 from ~\citet{TOD_2016}), an event in March-April 2010 (event 3), an event in April-May 2011 in the northern part of the region studied (events 6 and 7), an event propagating south-to-north in August-September and September-October 2011 (events 8 and 9), and an event in December 2011 (event 10). Although ~\citet{TOD_2016} only detected this last event for GPS station GISB, it seems that this event may have also extended farther to the north and the south. We then clearly see an event in the northern part of the region studied in August 2012 (event 12), an event in December 2012-January 2013 (event 13), an event in the southern part of the region studied in February-March 2013 (event 14), an event propagating from south to north in June-July and July-August 2013 (events 15 and 16), an event in September 2014 (events 20 and 21), an event in the southern part of the region studied in December 2014-January 2015 (events 22 and 23), and an event in June-July 2015 in the northern part of the region studied (event 26). It is unclear if the event near station ANAU in early 2010 (event 1) is visible in the wavelet details as it is too close to the beginning of the time series. The June-July 2010 event (event 4), the August 2010 event (event 5), and the March 2012 event (event 11), are not clearly visible in the wavelet details. The events in September-October 2013 (event 17), December 2013 (event 18), May-June 2014 (event 19), January-February (event 24) and February 2015 (events 25) are not clearly seen in the wavelet details, but there could be a small negative peak followed by a small positive peak at these times. Additionally, there could be two other events that are not in \citep{TOD_2016} in Fall 2010 (southern part of the region studied) and in Fall 2015. \\

Our wavelet-based method thus works well to detect transients in GPS data that could be slow slip events, even in the absence of tremor data. The choice of the appropriate threshold to decide that there is a transient and the levels of the wavelet details that we look at for the detection may still not be easily made. There is a difference between Cascadia and New Zealand in terms of which wavelet details to stack. In particular, as there is more time between two slow slip events in New Zealand than in Cascadia, the biggest slow slip events (early 2010, late 2011, 2013 and late 2014) can also be seen on the 9th level detail for New Zealand, whereas they could not be seen for Cascadia. We then use the method to detect slow slip events during the period 2016-2022, which was not covered by ~\citet{TOD_2016} (Figure 14). We note four large transients that could be slow slip events in late 2016, late 2017, early 2019 and mid-2021. There are also possible smaller events in the northern part of the area in mid-2018 and in most of the area studied in early 2020. Table 6 summarizes the slow slip events detected with the wavelet method for 2016-2022. \\

The method is thus applicable in regions where tremor data are not usable. To determine which wavelet levels to stack, we recommend analyzing each level detail. Look for spatially coherent patterns, wavelet details with energy at similar times and high signal-to-noise ratios. Look for alternating positive and negative peaks that are consistent with the expected direction of slow slip. Consider wavelet details with time scales ranging from the expected duration of slow slip events to the expected recurrence times between slow slip events. For Cascadia and New Zealand this would be weeks to years. Determination of a threshold is subjective. At large thresholds the large slow slip events should be clear. At smaller thresholds there is the possibility of identifying smaller events, but at the risk of false detections.

\section{Conclusion}

In this paper, we develop and test a new approach for detecting transient events in GPS time series, such as slow slip events. We used wavelet methods to analyze GNSS time series and tremor recordings of slow slip events in Cascadia, and GNSS time series in New Zealand. We used detrended GNSS data, applied the MODWT transform, and stacked the wavelet details over several nearby GNSS stations. As an independent check on the timing of slow slip events, we also computed the cumulative number of tremor in the vicinity of the GNSS stations, detrended this signal, and applied the MODWT transform. In both time series, we could then see simultaneous waveforms whose timing corresponds to the timing of slow slip events. We assumed that there is a slow slip event whenever the wavelet signal gets above a threshold. We verified that there is a good agreement between slow slip events detected with only GNSS data, and slow slip events detected with only tremor data. The wavelet-based detection method detects all events of magnitude higher than 6 as determined by independent event catalogs (e.g. ~\citep{MIC_2019}). We detected signals in the GPS data that could be magnitude 5 events, but it is not easy to differentiate between robust detections and false detections. We then applied the method to GNSS data in New Zealand and detected slow slip events consistent with the events previously detected by ~\citet{TOD_2016}.

\section*{Data and Resources}

The GPS recordings used for this analysis can be downloaded from the PANGA website ~\citep{PANGA} \url{http://www.panga.cwu.edu/} and the Geonet website \url{https://www.geonet.org.nz/}. The Python scripts used to analyze the data and make the figures can be found on the first author's Github account \url{https://github.com/ArianeDucellier/slowslip}. Figures 3 and 12 were created using GMT ~\citep{WES_1991}. Supplemental Material for this article includes three figures showing the effects of boundary conditions, missing data and common modes, and a figure showing four additional small displacements detected in the GPS data.

\section*{Acknowledgements}

The authors would like to thank two anonymous reviewers, the Associate Editor Jeanne Hardebeck and the Editor-in-Chief P. Martin Mai, whose comments greatly helped improve the manuscript. This work was funded by the grant from the National Science Foundation EAR-1358512. A.D. would like to thank Professor Donald Percival for introducing her to wavelet methods during his excellent class on Wavelets: Data Analysis, Algorithms and Theory taught at University of Washington.   

\section*{Declaration of Competing Interests}

The authors declare no competing interests.

\bibliographystyle{plainnat_modif}
\bibliography{bibliography}

\begin{thebibliography}{40}
\providecommand{\natexlab}[1]{#1}
\providecommand{\url}[1]{\texttt{#1}}
\expandafter\ifx\csname urlstyle\endcsname\relax
  \providecommand{\doi}[1]{doi: #1}\else
  \providecommand{\doi}{doi: \begingroup \urlstyle{rm}\Url}\fi

\bibitem[Aguiar et~al.(2009)Aguiar, A., Melbourne, T., and Scrivner,
  C.]{AGU_2009}
Aguiar, A., Melbourne, T., and Scrivner, C.
\newblock Moment release rate of {Cascadia} tremor constrained by {GPS}.
\newblock \emph{J. Geophys. Res.}, 114:\penalty0 B00A05, 2009.

\bibitem[Alba et~al.(2019)Alba, S., Weldon, R.~J., Livelybrooks, D., and
  Schmidt, D.~A.]{ALB_2019}
Alba, S., Weldon, R.~J., Livelybrooks, D., and Schmidt, D.~A.
\newblock Cascadia {ETS} events seen in tidal records (1980--2011).
\newblock \emph{Bull. Seismol. Soc. Am.}, 109\penalty0 (2):\penalty0 812--821,
  2019.

\bibitem[Audet and Kim(2016)Audet, P. and Kim, Y.]{AUD_2016}
Audet, P. and Kim, Y.
\newblock Teleseismic constraints on the geological environment of deep
  episodic slow earthquakes in subduction zone forearcs: A review.
\newblock \emph{Tectonophysics}, 670:\penalty0 1--15, 2016.

\bibitem[Bartlow(2020)Bartlow, N.~M.]{BAR_2020}
Bartlow, N.~M.
\newblock A long‐term view of episodic tremor and slip in {Cascadia}.
\newblock \emph{Geophysical Research Letters}, 43\penalty0 (3):\penalty0
  e2019GL085303, 2020.

\bibitem[Beroza and Ide(2011)Beroza, G. and Ide, S.]{BER_2011}
Beroza, G. and Ide, S.
\newblock Slow earthquakes and nonvolcanic tremor.
\newblock \emph{Annu. Rev. Earth Planet. Sci.}, 39:\penalty0 271--296, 2011.

\bibitem[Delahaye et~al.(2009)Delahaye, E., Townend, J., Reyners, M., and
  Rogers, G.]{DEL_2009}
Delahaye, E., Townend, J., Reyners, M., and Rogers, G.
\newblock Microseismicity but no tremor accompanying slow slip in the
  {Hikurangi} subduction zone, {New} {Zealand}.
\newblock \emph{Earth and Planetary Science Letters}, 277:\penalty0 21--28,
  2009.

\bibitem[Frank(2016)Frank, W.]{FRA_2016}
Frank, W.
\newblock Slow slip hidden in the noise: The intermittence of tectonic release.
\newblock \emph{Geophys. Res. Lett.}, 43:\penalty0 10125--10133, 2016.

\bibitem[{GPS/GNSS Network and Geodesy Laboratory: Central Washington
  University, other/seismic network}(1996)]{PANGA}
{GPS/GNSS Network and Geodesy Laboratory: Central Washington University,
  other/seismic network}.
\newblock {Pacific Northwest Geodetic Array (PANGA)}, 1996.
\newblock URL \url{http://www.panga.cwu.edu/}.

\bibitem[Hall et~al.(2018)Hall, K., Houston, H., and Schmidt, D.]{HAL_2018}
Hall, K., Houston, H., and Schmidt, D.
\newblock Spatial comparisons of tremor and slow slip as a constraint on fault
  strength in the northern {Cascadia} subduction zone.
\newblock \emph{Geochemistry, Geophysics, Geosystems}, 19\penalty0
  (8):\penalty0 2706--2718, 2018.

\bibitem[Hawthorne and Rubin(2013)Hawthorne, J.~C. and Rubin, A.~M.]{HAW_2013}
Hawthorne, J.~C. and Rubin, A.~M.
\newblock Short‐time scale correlation between slow slip and tremor in
  {Cascadia}.
\newblock \emph{Journal of Geophysical Research: Solid Earth}, 118:\penalty0
  1316--1329, 2013.

\bibitem[Hiramatsu et~al.(2008)Hiramatsu, Y., Watanabe, T., and Obara,
  K.]{HIR_2008}
Hiramatsu, Y., Watanabe, T., and Obara, K.
\newblock Deep low‐frequency tremors as a proxy for slip monitoring at plate
  interface.
\newblock \emph{Geophysical Research Letters}, 35:\penalty0 L13304, 2008.

\bibitem[Ide(2012)Ide, S.]{IDE_2012}
Ide, S.
\newblock Variety and spatial heterogeneity of tectonic tremor worldwide.
\newblock \emph{Journal of Geophysical Research}, 117:\penalty0 B03302, 2012.

\bibitem[Jiang et~al.(2012)Jiang, Y., Wdowinski, S., Dixon, T.~H., Hackl, M.,
  Protti, M., and Gonzalez, V.]{JIA_2012}
Jiang, Y., Wdowinski, S., Dixon, T.~H., Hackl, M., Protti, M., and Gonzalez, V.
\newblock Slow slip events in {Costa} {Rica} detected by continuous {GPS}
  observations, 2002-2011.
\newblock \emph{Geochemistry, Geophysics, Geosystems}, 13:\penalty0 Q04006,
  2012.

\bibitem[Kim et~al.(2011)Kim, M., Schwartz, S., and Bannister, S.]{KIM_2011}
Kim, M., Schwartz, S., and Bannister, S.
\newblock Non‐volcanic tremor associated with the {March} 2010 {Gisborne}
  slow slip event at the {Hikurangi} subduction margin, {New} {Zealand}.
\newblock \emph{Geophysical Research Letters}, 38:\penalty0 L14301, 2011.

\bibitem[Kumar and Foufoula-Georgiou(1997)Kumar, P. and Foufoula-Georgiou,
  E.]{KUM_1997}
Kumar, P. and Foufoula-Georgiou, E.
\newblock Wavelet analysis for geophysical applications.
\newblock \emph{Rev. Geophys.}, 35\penalty0 (4):\penalty0 385--412, 1997.

\bibitem[Li et~al.(2016)Li, S., Freymueller, J., and McCaffrey, R.]{LI_2016}
Li, S., Freymueller, J., and McCaffrey, R.
\newblock Slow slip events and time‐dependent variations in locking beneath
  {Lower} {Cook} {Inlet} of the {Alaska}‐{Aleutian} subduction zone.
\newblock \emph{Journal of Geophysical Research: Solid Earth}, 121:\penalty0
  1060--1079, 2016.

\bibitem[Michel et~al.(2019)Michel, S., Gualandi, A., and Avouac,
  J.-P.]{MIC_2019}
Michel, S., Gualandi, A., and Avouac, J.-P.
\newblock Interseismic coupling and slow slip events on the {Cascadia}
  megathrust.
\newblock \emph{Pure Appl. Geophys.}, 176:\penalty0 3867--3891, 2019.

\bibitem[Nishimura et~al.(2013)Nishimura, T., Matsuzawa, T., and Obara,
  K.]{NIS_2013}
Nishimura, T., Matsuzawa, T., and Obara, K.
\newblock Detection of short‐term slow slip events along the {Nankai}
  {Trough}, southwest {Japan}, using {GNSS} data.
\newblock \emph{Journal of Geophysical Research: Solid Earth}, 118:\penalty0
  3112--3125, 2013.

\bibitem[Nuyen and Schmidt(2021)Nuyen, C.~P. and Schmidt, D.~A.]{NUY_2021}
Nuyen, C.~P. and Schmidt, D.~A.
\newblock Filling the gap in {Cascadia}: The emergence of low‐amplitude
  long‐term slow slip.
\newblock \emph{Geochemistry, Geophysics, Geosystems}, 22\penalty0
  (3):\penalty0 e2020GC009477, 2021.

\bibitem[Obara et~al.(2004)Obara, K., Hirose, H., Yamamizu, F., and Kasahara,
  K.]{OBA_2004}
Obara, K., Hirose, H., Yamamizu, F., and Kasahara, K.
\newblock Episodic slow slip events accompanied by non-volcanic tremors in
  southwest {Japan} subduction zone.
\newblock \emph{Geophysical Research Letters}, 31:\penalty0 L23602, 2004.

\bibitem[Ohtani et~al.(2010)Ohtani, R., McGuire, J., and Segall, P.]{OHT_2010}
Ohtani, R., McGuire, J., and Segall, P.
\newblock Network strain filter: A new tool for monitoring and detecting
  transient deformation signals in {GPS} arrays.
\newblock \emph{J. Geophys. Res.}, 115:\penalty0 B12418, 2010.

\bibitem[Percival and Walden(2000)Percival, D. and Walden, A.]{PER_2000}
Percival, D. and Walden, A.
\newblock \emph{Wavelet {Methods} for {Time} {Series} {Analysis}}.
\newblock Cambridge {Series} in {Statistical} and {Probabilistic}
  {Mathematics}. Cambridge {University} {Press}, New York, NY, USA, 2000.

\bibitem[Preston et~al.(2003)Preston, L., Creager, K., Crosson, R., Brocher,
  T., and Trehu, A.]{PRE_2003}
Preston, L., Creager, K., Crosson, R., Brocher, T., and Trehu, A.
\newblock Intraslab earthquakes: Dehydration of the {Cascadia slab}.
\newblock \emph{Science}, 302:\penalty0 1197--1200, 2003.

\bibitem[Radiguet et~al.(2012)Radiguet, M., Cotton, F., Vergnolle, M.,
  Campillo, M., Walpersdorf, A., Cotte, N., and Kostoglodov, V.]{RAD_2012}
Radiguet, M., Cotton, F., Vergnolle, M., Campillo, M., Walpersdorf, A., Cotte,
  N., and Kostoglodov, V.
\newblock Slow slip events and strain accumulation in the {Guerrero} gap,
  {Mexico}.
\newblock \emph{Journal of Geophysical Research: Solid Earth}, 117:\penalty0
  B04305, 2012.

\bibitem[Rogers and Dragert(2003)Rogers, G. and Dragert, H.]{ROG_2003}
Rogers, G. and Dragert, H.
\newblock Tremor and slip on the {Cascadia} subduction zone: The chatter of
  silent slip.
\newblock \emph{Science}, 300\penalty0 (5627):\penalty0 1942--1943, 2003.

\bibitem[Schmidt and Gao(2010)Schmidt, D.~A. and Gao, H.]{SCH_2010}
Schmidt, D.~A. and Gao, H.
\newblock Source parameters and time‐dependent slip distributions of slow
  slip events on the {Cascadia} subduction zone from 1998 to 2008.
\newblock \emph{Journal of Geophysical Research: Solid Earth}, 115:\penalty0
  B00A18, 2010.

\bibitem[Shelly et~al.(2007)Shelly, D., Beroza, G., and Ide,
  S.]{SHE_2007_nature}
Shelly, D., Beroza, G., and Ide, S.
\newblock Non-volcanic tremor and low-frequency earthquake swarms.
\newblock \emph{Nature}, 446:\penalty0 305--307, 2007.

\bibitem[Szeliga et~al.(2004)Szeliga, W., Melbourne, T., Miller, M., and
  Santillan, V.]{SZE_2004}
Szeliga, W., Melbourne, T., Miller, M., and Santillan, V.
\newblock Southern {Cascadia} episodic slow earthquakes.
\newblock \emph{Geophys. Res. Lett.}, 31:\penalty0 L16602, 2004.

\bibitem[Szeliga et~al.(2008)Szeliga, W., Melbourne, T., Santillan, M., and
  Miller, M.]{SZE_2008}
Szeliga, W., Melbourne, T., Santillan, M., and Miller, M.
\newblock {GPS} constraints on 34 slow slip events within the {Cascadia}
  subduction zone, 1997-2005.
\newblock \emph{J. Geophys. Res.}, 113:\penalty0 B04404, 2008.

\bibitem[Todd and Schwartz(2016)Todd, E. and Schwartz, S.]{TOD_2016}
Todd, E. and Schwartz, S.
\newblock Tectonic tremor along the northern {Hikurangi} {Margin}, {New}
  {Zealand}, between 2010 and 2015.
\newblock \emph{J. Geophys. Res. Solid Earth}, 121:\penalty0 8706--8719, 2016.

\bibitem[Vergnolle et~al.(2010)Vergnolle, M., Walpersdorf, A., Kostoglodov, V.,
  Tregoning, P., Santiago, J.~A., Cotte, N., and Franco, S.~I.]{VER_2010}
Vergnolle, M., Walpersdorf, A., Kostoglodov, V., Tregoning, P., Santiago,
  J.~A., Cotte, N., and Franco, S.~I.
\newblock Slow slip events in {Mexico} revised from the processing of 11 year
  {GPS} observations.
\newblock \emph{Journal of Geophysical Research: Solid Earth}, 115:\penalty0
  B08403, 2010.

\bibitem[Wallace(2020)Wallace, L.~M.]{WAL_2020}
Wallace, L.~M.
\newblock Slow slip events in {New} {Zealand}.
\newblock \emph{Annual Review of Earth and Planetary Sciences}, 48:\penalty0
  175--203, 2020.

\bibitem[Wallace et~al.(2012)Wallace, L.~M., Beavan, J., Bannister, S., and
  Williams, C.]{WAL_2012}
Wallace, L.~M., Beavan, J., Bannister, S., and Williams, C.
\newblock Simultaneous long‐term and short‐term slow slip events at the
  {Hikurangi} subduction margin, {New} {Zealand}: Implications for processes
  that control slow slip event occurrence, duration, and migration.
\newblock \emph{Journal of Geophysical Research: Solid Earth}, 117:\penalty0
  B11402, 2012.

\bibitem[Wallace and Beavan(2010)Wallace, L. and Beavan, J.]{WAL_2010}
Wallace, L. and Beavan, J.
\newblock Diverse slow slip behavior at the {Hikurangi} subduction margin,
  {New} {Zealand}.
\newblock \emph{Journal of Geophysical Research}, 115:\penalty0 B12402, 2010.

\bibitem[Wallace and Eberhart-Phillips(2013)Wallace, L. and Eberhart-Phillips,
  D.]{WAL_2013}
Wallace, L. and Eberhart-Phillips, D.
\newblock Newly observed, deep slow slip events at the central {Hikurangi}
  margin, {New} {Zealand}: Implications for downdip variability of slow slip
  and tremor, and relationship to seismic structure.
\newblock \emph{Geophysical Research Letters}, 40:\penalty0 5393--5398, 2013.

\bibitem[Wech(2010)Wech, A.]{WEC_2010}
Wech, A.
\newblock Interactive tremor monitoring.
\newblock \emph{Seismol. Res. Lett.}, 81\penalty0 (4):\penalty0 664--669, 2010.

\bibitem[Wech(2016)Wech, A.]{WEC_2016}
Wech, A.
\newblock Extending {Alaska}'s plate boundary; tectonic tremor generated by
  {Yakutat} subduction.
\newblock \emph{Geology}, 44\penalty0 (7):\penalty0 587--590, 2016.

\bibitem[Wei et~al.(2012)Wei, M., McGuire, J., and Richardson, E.]{WEI_2012}
Wei, M., McGuire, J., and Richardson, E.
\newblock A slow slip event in the south central {Alaska} {Subduction} {Zone}.
\newblock \emph{Geophys. Res. Lett.}, 39:\penalty0 L15309, 2012.

\bibitem[Wessel and Smith(1991)Wessel, P. and Smith, W. H.~F.]{WES_1991}
Wessel, P. and Smith, W. H.~F.
\newblock Free software helps map and display data.
\newblock \emph{EOS Trans. AGU}, 72:\penalty0 441, 1991.

\bibitem[Williams et~al.(2013)Williams, C.~A., Eberhart-Phillips, D.,
  Bannister, S., Barker, D.~H., Henrys, S., Reyners, M., and Sutherland,
  R.]{WIL_2013}
Williams, C.~A., Eberhart-Phillips, D., Bannister, S., Barker, D.~H., Henrys,
  S., Reyners, M., and Sutherland, R.
\newblock Revised interface geometry for the {Hikurangi} subduction zone, {New}
  {Zealand}.
\newblock \emph{Seismological Research Letters}, 84\penalty0 (6):\penalty0
  1066--1073, 2013.

\end{thebibliography}

\newpage

\section*{Addresses}

Ariane Ducellier. University of Washington, Department of Earth and Space Sciences, Box 351310, 4000 15th Avenue NE Seattle, WA 98195-1310. \href{mailto:ariane.ducellier.pro@gmail.com}{ariane.ducellier.pro@gmail.com}\\

Kenneth C. Creager. University of Washington, Department of Earth and Space Sciences, Box 351310, 4000 15th Avenue NE Seattle, WA 98195-1310. \\

David A. Schmidt. University of Washington, Department of Earth and Space Sciences, Box 351310, 4000 15th Avenue NE Seattle, WA 98195-1310.

\newpage

\section*{Tables}

\begin{table}[hbt!]
\caption{Episodic Tremor and Slip events with M $>$ 6 identified by MODWT in both the GPS and the tremor data. The duration and the number of tremor are from the tremor catalog of the PNSN. The event number and the magnitude are from the slow slip catalog of ~\citet{MIC_2019}.}
 \centering
 \begin{tabular}{c c c c c}
 \hline
 Time & Duration (days) & Number of tremor (hours) & Event number & Magnitude \\
 \hline
 2007.06 & 28 & 398 & 3 & 6.68 \\
 2008.36 & 25 & 402 & 10 & 6.56 \\
 2009.35 & 24 & 248 & 16 & 6.49 \\
 2010.63 & 29 & 518 & 24 & 6.54 \\
 2011.60 & 37 & 479 & 30 & 6.47 \\
 2012.72 & 37 & 620 & 34 & 6.54 \\
 2013.71 & 27 & 423 & 41 & 6.58 \\
 2014.65 & 15 & 190 & 48 & 6.03 \\
 2014.89 & 38 & 385 & 51 & 6.40 \\
 2016.11 & 43 & 421 & 54 & 6.79 \\
 2017.23 & 19 & 279 & 59 & 6.61 \\
 2018.49 & 22 & 381 & & \\
 2019.23 & 34 & 195 & & \\
 2019.88 & 16 & 205 & & \\
 2020.79 & 26 & 193 & & \\
 2020.86 & 12 & 162 & & \\
 2021.09 & 14 & 230 & & \\
 \hline
 \end{tabular}
 \end{table}

 \begin{table}[hbt!]
 \caption{Magnitude 5 to 6 events from ~\citet{MIC_2019}.}
 \centering
 \begin{tabular}{c c c c}
 \hline
 Time & Event number & Magnitude & Sub-event of bigger event \\
 \hline
 2007.06 & 1 & 5.64 & Yes \\
 2007.08 & 2 & 5.91 & Yes \\
 2008.38 & 11 & 5.50 & Yes \\
 2009.16 & 14 & 5.50 & No \\
 2009.36 & 17 & 5.32 & Yes \\
 2010.63 & 25 & 5.76 & Yes \\
 2011.66 & 31 & 5.61 & Yes \\
 2011.66 & 32 & 5.32 & Yes \\
 2012.69 & 35 & 5.56 & Yes \\
 2013.74 & 42 & 5.71 & Yes \\
 2014.69 & 49 & 5.31 & Yes \\
 2014.93 & 52 & 5.39 & Yes \\
 2016.03 & 57 & 5.80 & Yes \\
 2017.13 & 60 & 5.43 & Yes \\
 2017.22 & 61 & 5.37 & Yes \\
 \hline
 \end{tabular}
 \end{table}

 \begin{table}[hbt!]
 \caption{Number of GPS stations used for the stacking for each location on Figure 3 and number of common stations with the location immediately to the north and the location immediately to the south.}
 \centering
 \begin{tabular}{c c c c c}
 \hline
 Index & Latitude & Number of stations & Common stations (north) & Common stations (south) \\
 \hline
0 & 47.2 & 15 & 14 & \\
1 & 47.3 & 18 & 17 & 14 \\
2 & 47.4 & 24 & 20 & 17 \\
3 & 47.5 & 21 & 20 & 20 \\
4 & 47.6 & 22 & 14 & 20 \\
5 & 47.7 & 17 & 12 & 14 \\
6 & 47.8 & 13 & 8 & 12 \\
7 & 47.9 & 10 & 9 & 8 \\
8 & 48.0 & 10 & 7 & 9 \\
9 & 48.1 & 8 & 7 & 7 \\
10 & 48.2 & 10 & 8 & 7 \\
11 & 48.3 & 9 & 9 & 8 \\
12 & 48.4 & 9 & 5 & 9 \\
13 & 48.5 & 7 & 5 & 5 \\
14 & 48.6 & 6 & 5 & 5 \\
15 & 48.7 & 5 & & 5 \\
 \hline
 \end{tabular}
 \end{table}

 \begin{table}[hbt!]
 \caption{Cascadia catalog of slow slip events based only on MODWT analysis of GPS time series and inferring that the transition of red followed immediate by blue marks a slow slip event. First four columns are the start and end times and start and end latitudes of the green bars in Figure 11. The fifth column is 1 for robust detection and 2 if not as robust. Column 6 is the time difference in years between the mid times of the GPS catalog and the nearest mid times of the tremor catalog summarized in Table 1.}
 \centering
 \begin{tabular}{c c c c c c}
 \hline
start time & end time & start latitude & end latitude & &dT tremor catalog \\
 \hline
2007.06 & 2007.10 & 47.16 & 48.72 & 1 & 0.02 \\
2008.30 & 2008.40 & 47.35 & 48.73 & 1 & 0.01 \\
2009.35 & 2009.44 & 47.92 & 48.73 & 1 & 0.05 \\
2010.12 & 2010.15 & 47.32 & 48.73 & 1 & 0.50 no match \\
2010.61 & 2010.64 & 47.17 & 48.72 & 1 & 0.00 \\
2011.57 & 2011.61 & 47.18 & 48.68 & 1 & 0.01 \\
2012.65 & 2012.65 & 48.74 & 47.76 & 1 & 0.05 \\
2013.71 & 2013.75 & 47.47 & 48.73 & 1 & 0.02 \\
2014.89 & 2014.90 & 48.73 & 47.79 & 1 & 0.01 \\
2015.98 & 2016.09 & 48.73 & 47.20 & 1 & 0.08 \\
2017.17 & 2017.24 & 47.38 & 48.72 & 1 & 0.02 \\
2018.35 & 2018.36 & 47.48 & 47.93 & 1 & 0.13 part of same event? \\
2018.48 & 2018.50 & 48.72 & 48.09 & 1 & 0.00 \\
2019.32 & 2019.34 & 47.17 & 47.72 & 2 & 0.10 \\
2019.90 & 2019.91 & 48.47 & 48.72 & 2 & 0.02 \\
2020.79 & 2020.83 & 47.18 & 48.13 & 1 & 0.02 \& 0.05 \\
2021.11 & 2021.12 & 48.75 & 48.48 & 2 & 0.02 \\
\hline
 \end{tabular}
 \end{table}

 \begin{table}[hbt!]
 \caption{New Zealand catalog of slow slip events for 2010-2016 based only on MODWT analysis of GPS time series and inferring that the transition of red followed immediate by blue marks a slow slip event. First four columns are the start and end times and start and end latitudes of the green bars in Figure 13. The fifth column is 1 for robust detection and 2 if not as robust.}
 \centering
 \begin{tabular}{c c c c c}
 \hline
start time & end time & start latitude & end latitude & \\
 \hline
2010.05 & 2010.07 & -39.67 & -39.12 & 1 \\
2010.19 & 2010.22 & -39.12 & -38.07 & 1 \\
2010.75 & 2010.76 & -39.73 & -39.41 & 1 \\
2011.36 & 2011.37 & -38.22 & -38.02 & 2 \\
2011.71 & 2011.74 & -37.97 & -38.41 & 1 \\
2011.67 & 2011.71 & -39.73 & -38.91 & 1 \\
2011.92 & 2011.95 & -38.84 & -38.16 & 1 \\
2012.63 & 2012.63 & -39.42 & -39.62 & 2 \\
2012.64 & 2012.66 & -38.53 & -38.02 & 1 \\
2012.95 & 2012.96 & -38.32 & -37.98 & 1 \\
2013.15 & 2013.16 & -38.87 & -39.72 & 1 \\
2013.55 & 2013.57 & -38.62 & -38.01 & 1 \\
2013.74 & 2013.74 & -38.77 & -38.97 & 2 \\
2013.92 & 2013.93 & -38.17 & -37.98 & 2 \\
2013.91 & 2013.95 & -39.37 & -39.73 & 1 \\
2014.78 & 2014.79 & -38.03 & -39.03 & 1 \\
2014.96 & 2015.00 & -39.07 & -39.72 & 1 \\
2015.53 & 2015.53 & -39.42 & -39.72 & 1 \\ 
2015.52 & 2015.55 & -37.97 & -38.43 & 1 \\
2015.78 & 2015.79 & -38.77 & -39.37 & 1 \\
\hline
 \end{tabular}
 \end{table}

 \begin{table}[hbt!]
 \caption{New Zealand catalog of slow slip events for 2016-2022 based only on MODWT analysis of GPS time series and inferring that the transition of red followed immediate by blue marks a slow slip event. First four columns are the start and end times and start and end latitudes of the green bars in Figure 13. The fifth column is 1 for robust detection and 2 if not as robust.}
 \centering
 \begin{tabular}{c c c c c}
 \hline
start time & end time & start latitude & end latitude & \\
 \hline
2016.84 & 2016.90 & -37.96 & -39.72 & 1 \\
2017.10 & 2017.10 & -38.78 & -39.00 & 2 \\
2017.73 & 2017.78 & -37.98 & -38.51 & 1 \\
2018.04 & 2018.06 & -38.58 & -39.07 & 1 \\
2018.63 & 2018.64 & -38.27 & -37.97 & 2 \\
2019.26 & 2019.33 & -37.97 & -39.73 & 1 \\
2020.09 & 2020.12 & -37.97 & -38.23 & 2 \\
2020.34 & 2020.35 & -37.96 & -39.72 & 1 \\
2020.33 & 2020.33 & -37.96 & -38.10 & 2 \\
2020.32 & 2020.32 & -38.62 & -38.79 & 2 \\
2020.36 & 2020.37 & -39.70 & -39.35 & 2 \\
2021.11 & 2021.11 & -39.51 & -39.64 & 2 \\
2021.39 & 2021.47 & -39.72 & -38.08 & 1 \\
 \hline
 \end{tabular}
 \end{table}

\newpage

\section*{Figures}

\begin{figure}
\noindent\includegraphics[width=11cm, trim={0cm 0cm 0cm 0cm},clip]{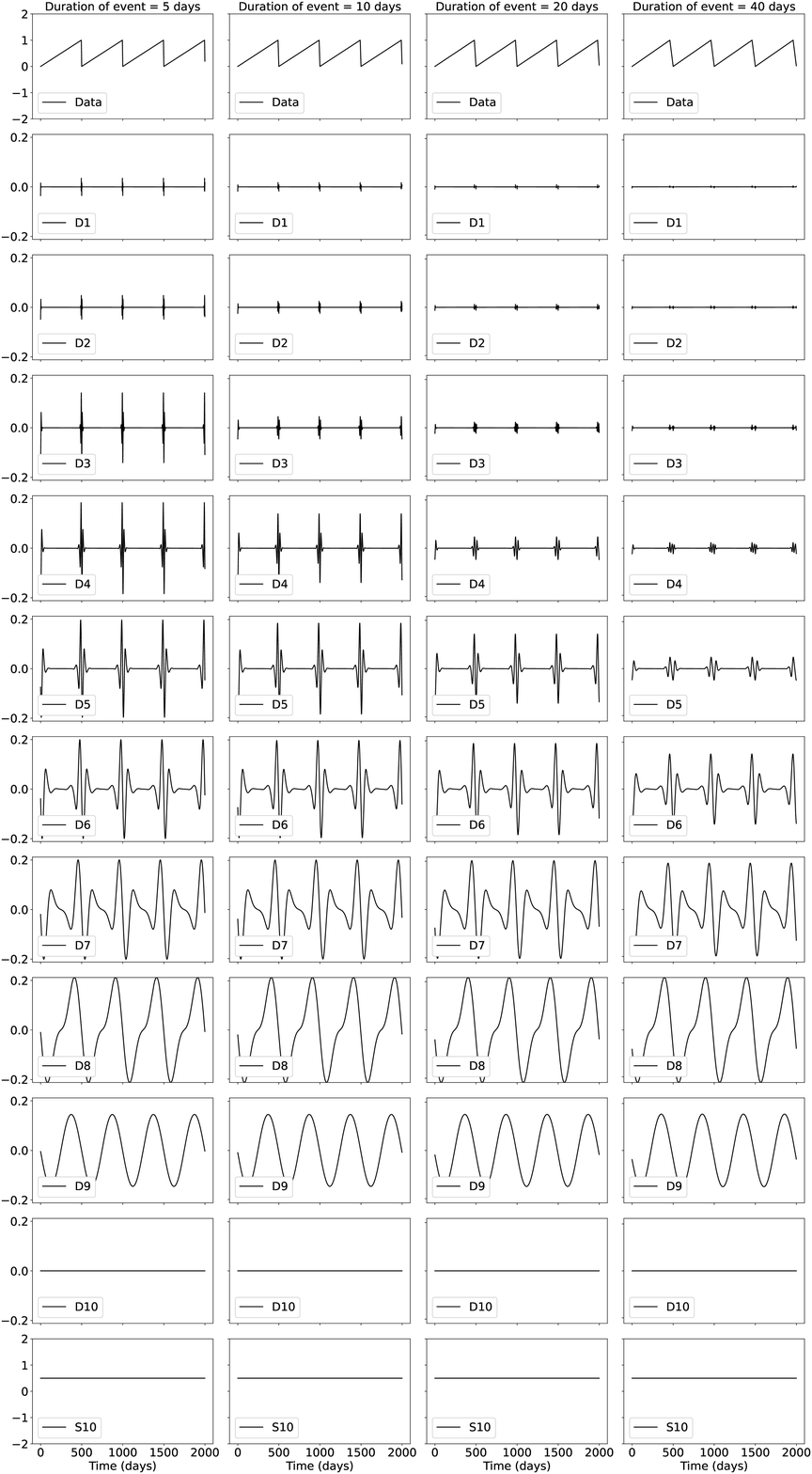}
\caption{Demonstration of a wavelet decomposition for a synthetic dataset. A synthetic time series is created (top row) with steps of period 500 days, and transient durations of 2 days (left), 5 days, 10 days, and 20 days (right). The resulting details and smooths are shown in increasing level. The amplitude of the synthetic time series is normalized to 1, and the details and smooths show the relative amplitude.}
\label{pngfiguresample}
\end{figure}

\begin{figure}
\noindent\includegraphics[width=\textwidth, trim={0cm 0cm 0cm 0cm},clip]{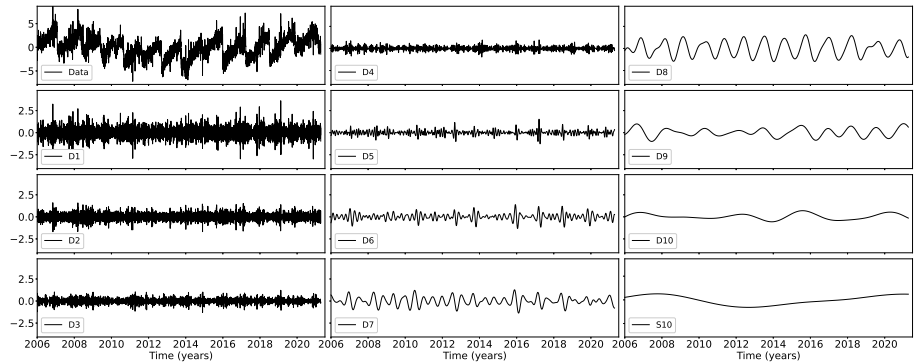}
\caption{Top left: East-west displacement recorded at GPS station PGC5. The resulting details and smooth of the wavelet decomposition are shown in increasing level from top to bottom and from left to right.}
\label{pngfiguresample}
\end{figure}

\begin{figure}
\noindent\includegraphics[width=\textwidth, trim={0cm 0cm 0cm 0cm},clip]{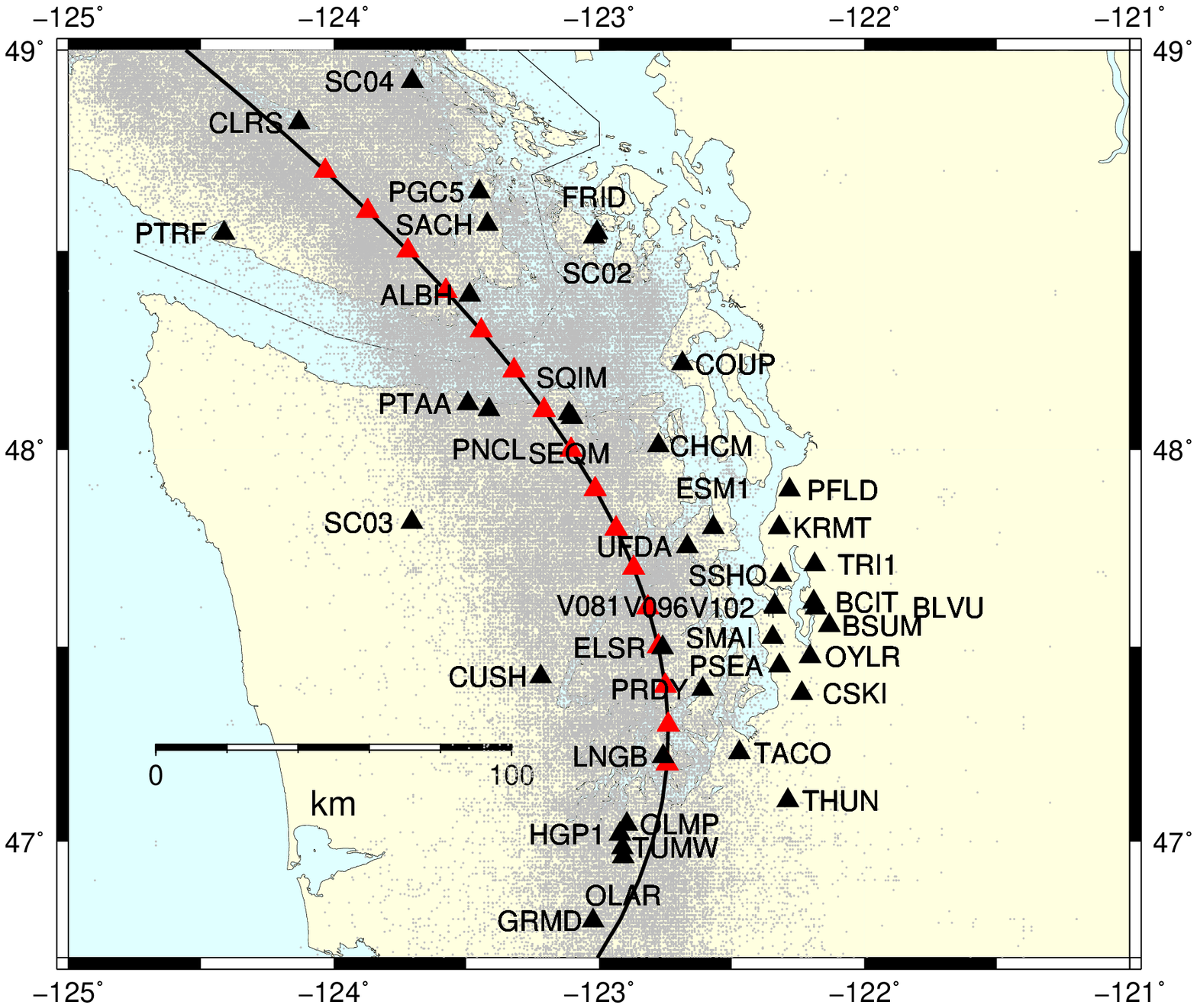}
\caption{GPS stations used in this study (black triangles). The black line represents the 40 km depth contour of the plate boundary model by ~\citet{PRE_2003}. The red triangles are the locations where we stack the GPS data. The small grey dots are all the tremor locations from the PNSN catalog.}
\label{pngfiguresample}
\end{figure}

\begin{figure}
\noindent\includegraphics[width=\textwidth, trim={0cm 0cm 0cm 0cm},clip]{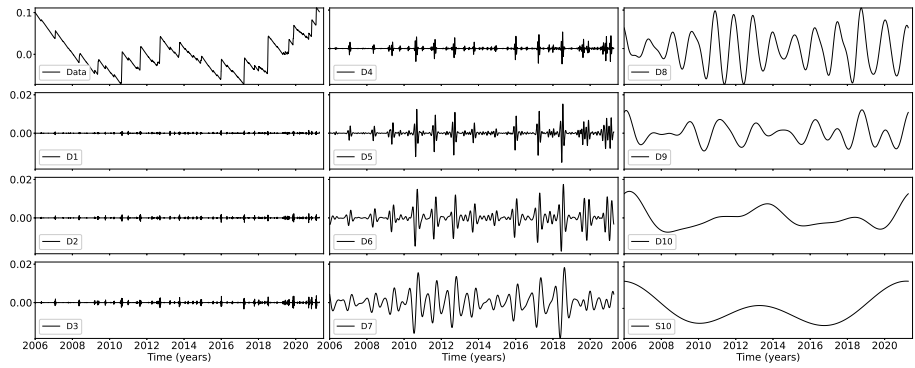}
\caption{Details and smooth of the wavelet decomposition of the detrended cumulative tremor count around the third northernmost red triangles on Figure 3 (latitude 48.5).}
\label{pngfiguresample}
\end{figure}

\begin{figure}
\noindent\includegraphics[width=\textwidth, trim={0cm 0cm 0cm 0cm},clip]{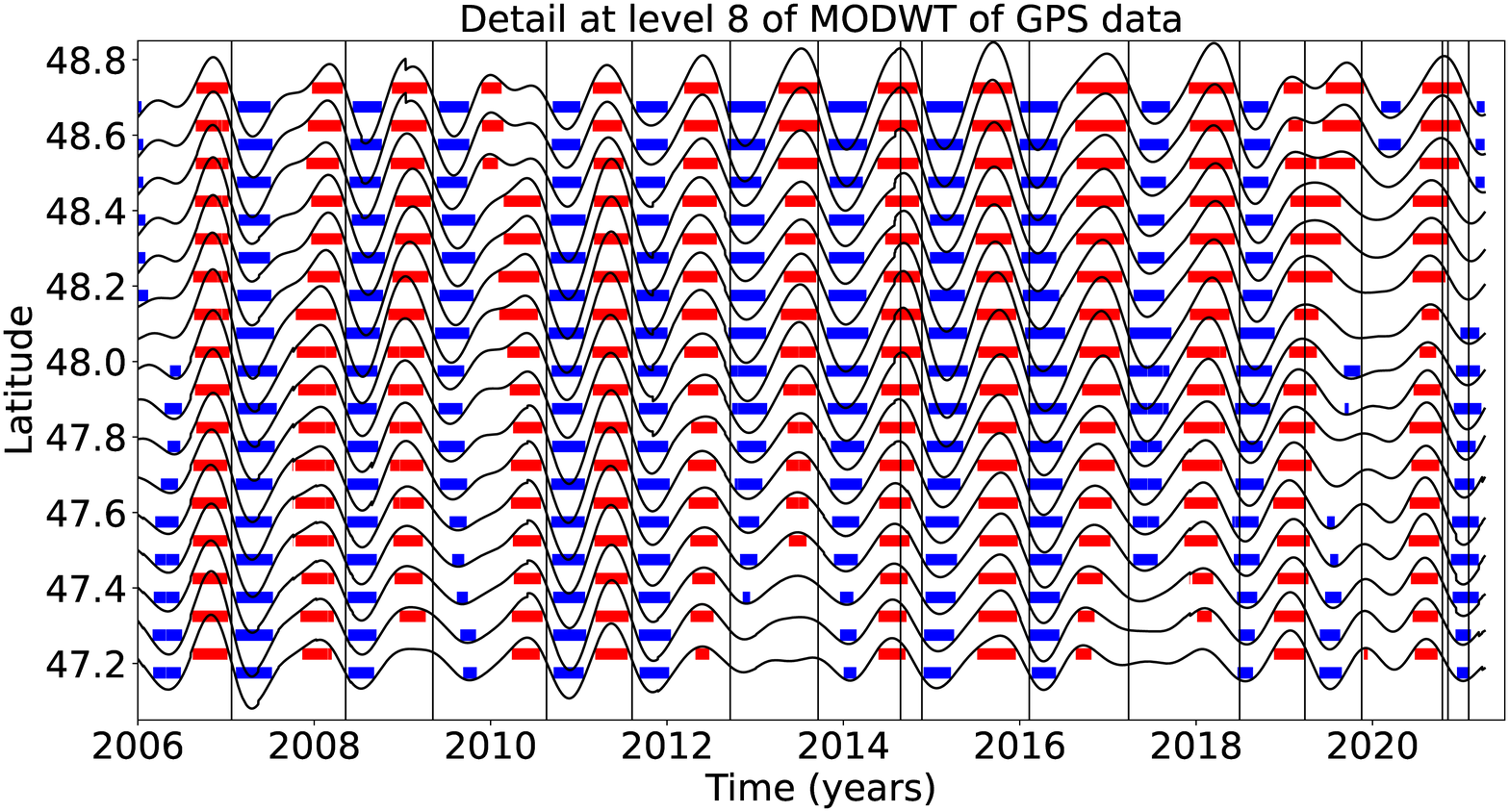}

\noindent\includegraphics[width=\textwidth, trim={0cm 0cm 0cm 0cm},clip]{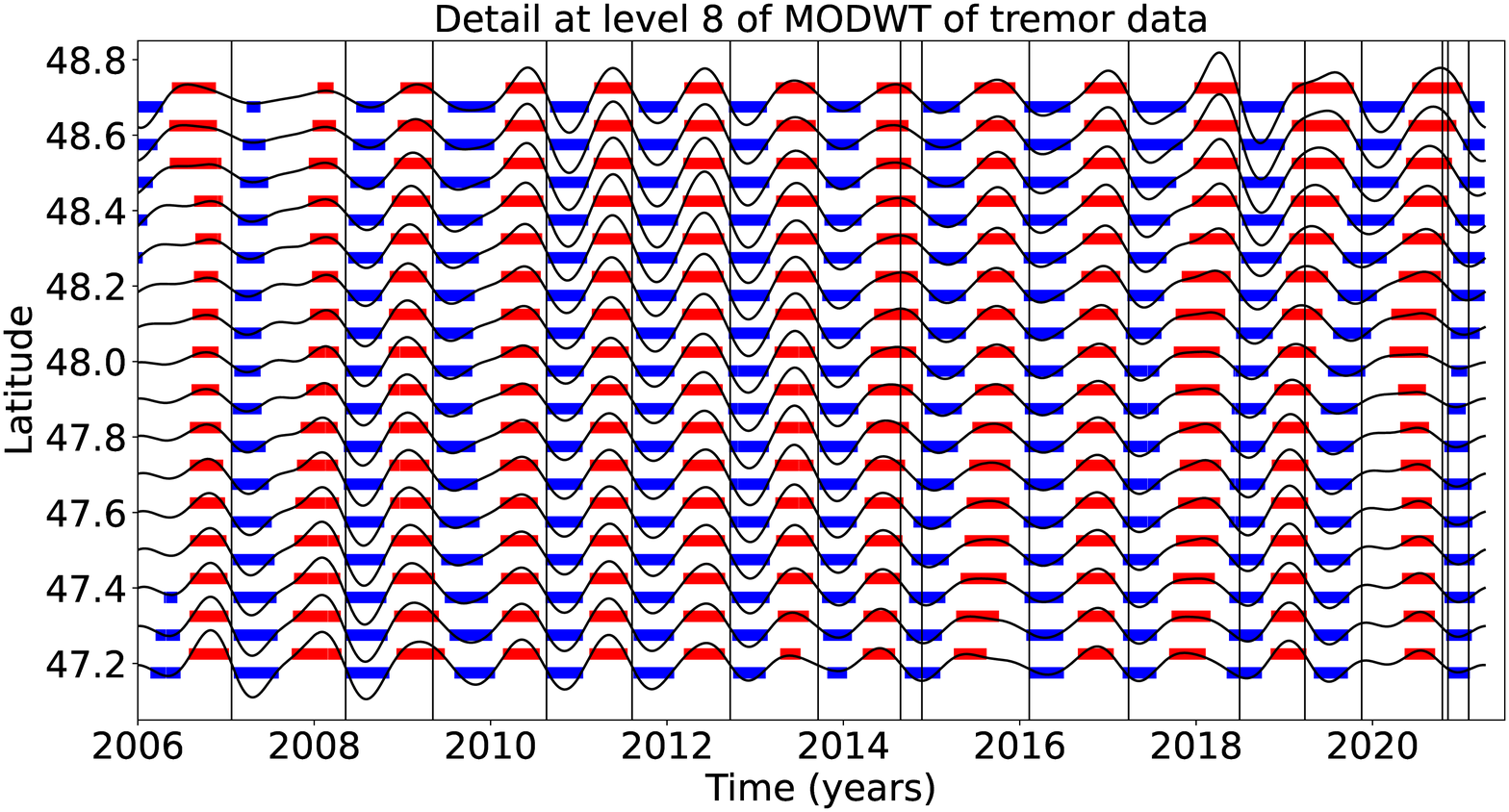}
\caption{Top: Stacked 8th level details of the wavelet decomposition of the displacement over all the GPS stations located in a 50 km radius of a given point, for the 16 red triangles indicated in Figure 3. Bottom: 8th level detail multiplied by -1 of the cumulative tremor count in a 50 km radius of a given point for the same 16 locations. The black lines represent the timings of the ETS events from Table 1. We mark by a red rectangle every time where the amplitude is higher than a threshold of 0.4 mm (for the GPS) or 0.003 (for the tremor, that is about 17 times the average value of the signal). We mark by a blue rectangle every time where the amplitude is lower than minus the threshold.}
\label{pngfiguresample}
\end{figure}

\begin{figure}
\noindent\includegraphics[width=\textwidth, trim={0cm 0cm 0cm 0cm},clip]{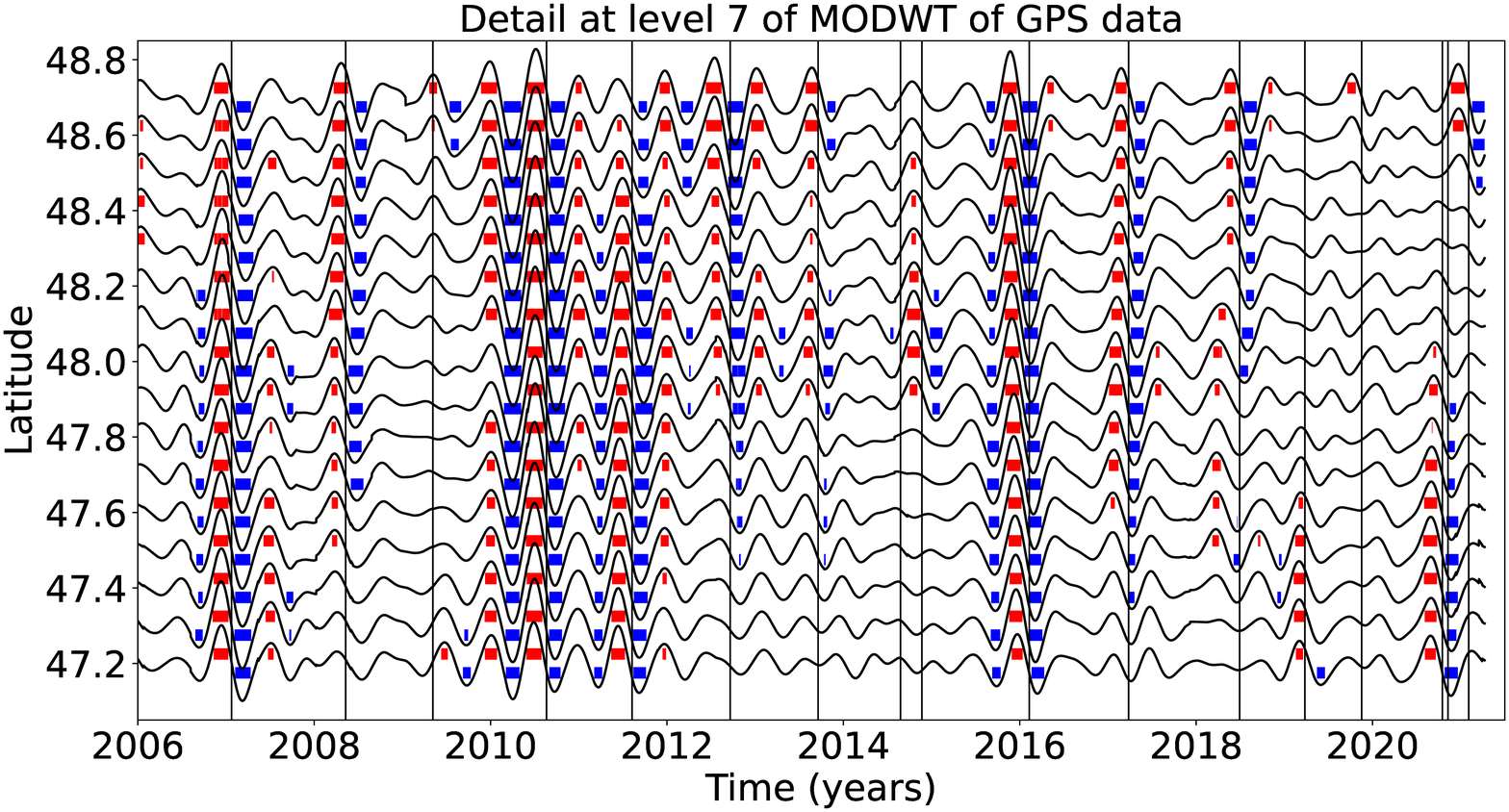}

\noindent\includegraphics[width=\textwidth, trim={0cm 0cm 0cm 0cm},clip]{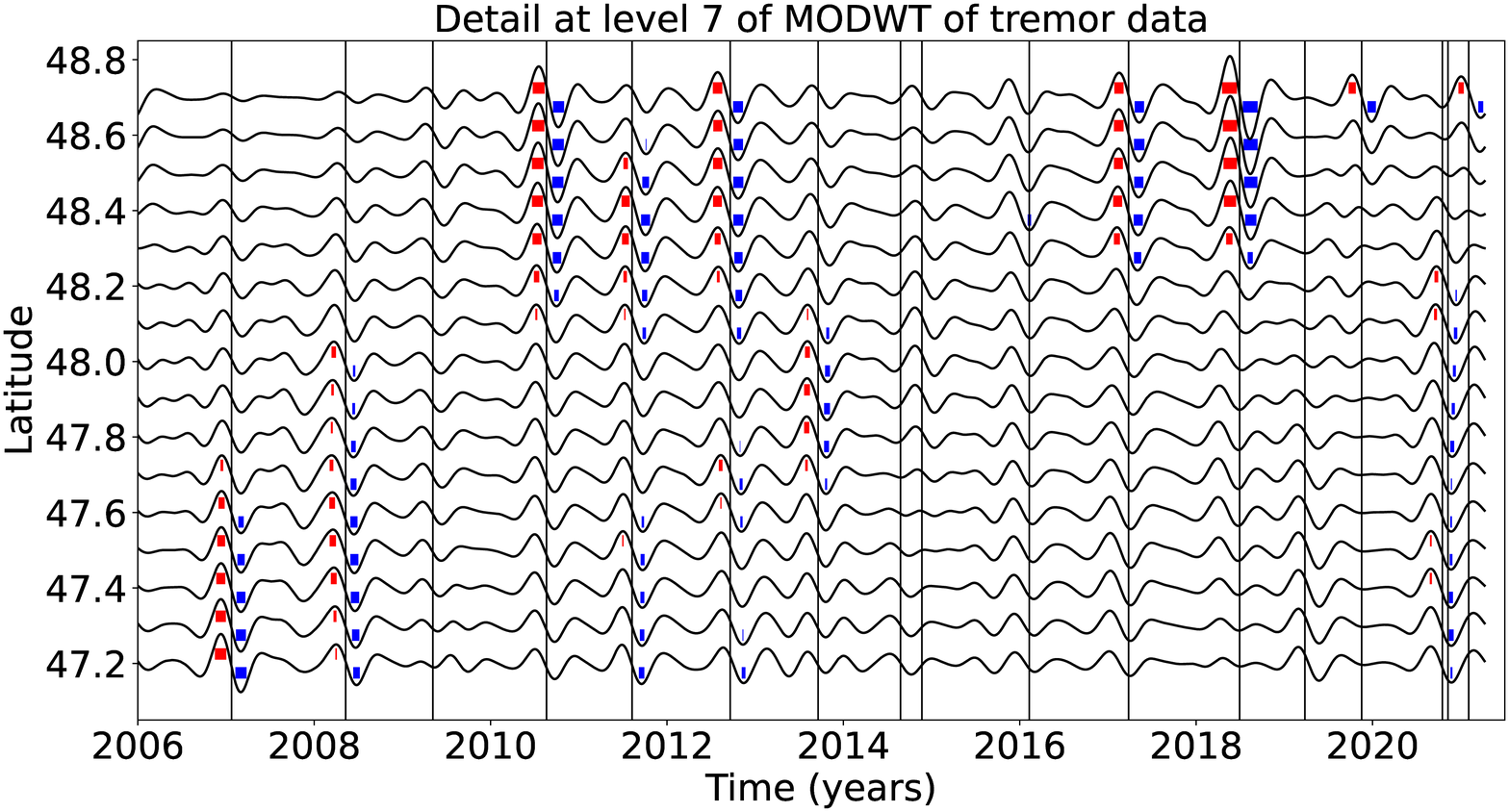}
\caption{Top: Stacked 7th level details of the wavelet decomposition of the displacement over all the GPS stations located in a 50 km radius of a given point, for the 16 red triangles indicated in Figure 3. Bottom: 7th level detail multiplied by -1 of the cumulative tremor count in a 50 km radius of a given point for the same 16 locations. The black lines represent the timings of the ETS events from Table 1. We mark by a red rectangle every time where the amplitude is higher than a threshold of 0.5 mm (for the GPS) or 0.01 (for the tremor, that is about 56 times the average value of the signal). We mark by a blue rectangle every time where the amplitude is lower than minus the threshold.}
\label{pngfiguresample}
\end{figure}

\begin{figure}
\noindent\includegraphics[width=\textwidth, trim={0cm 0cm 0cm 0cm},clip]{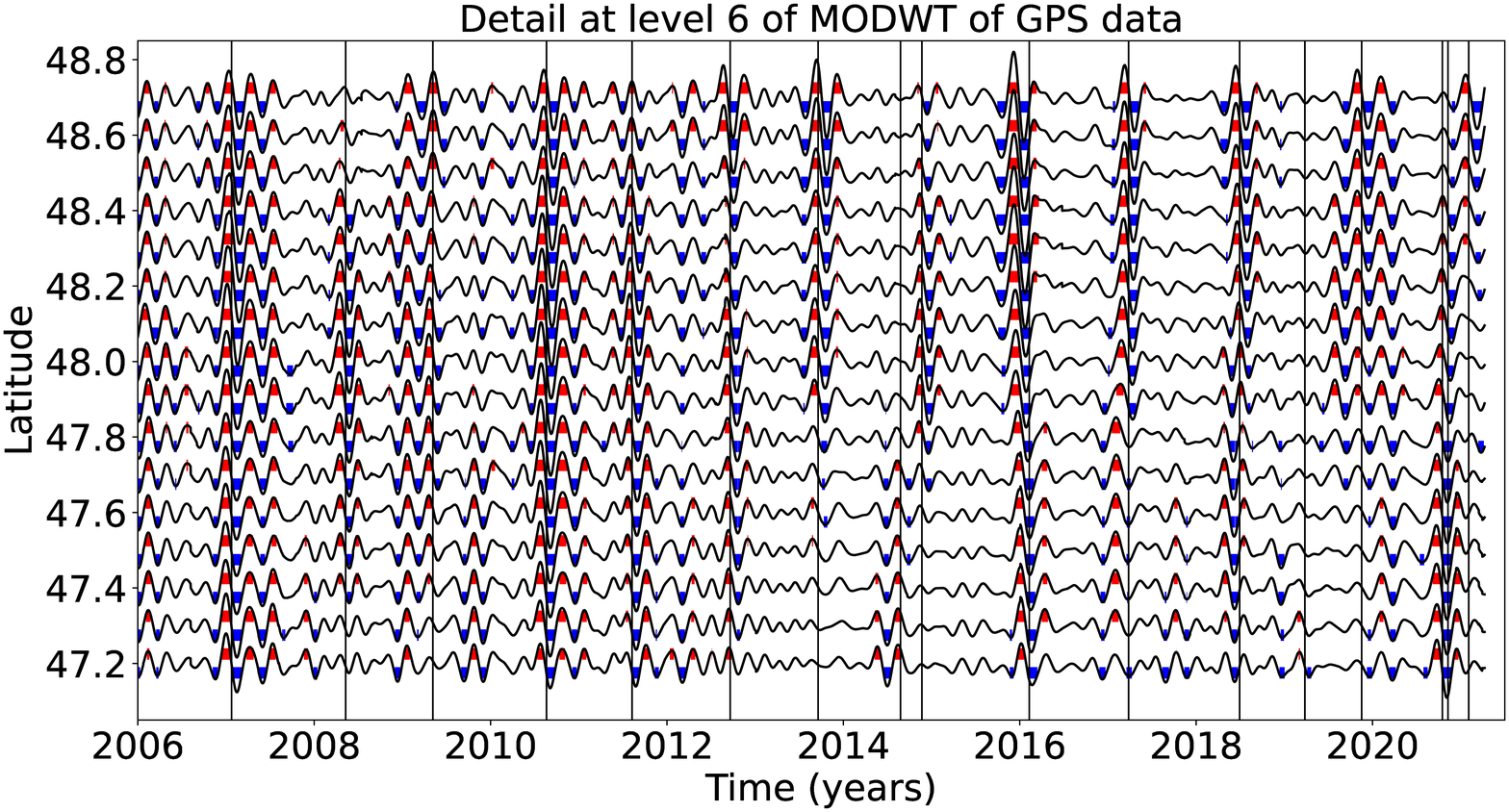}

\noindent\includegraphics[width=\textwidth, trim={0cm 0cm 0cm 0cm},clip]{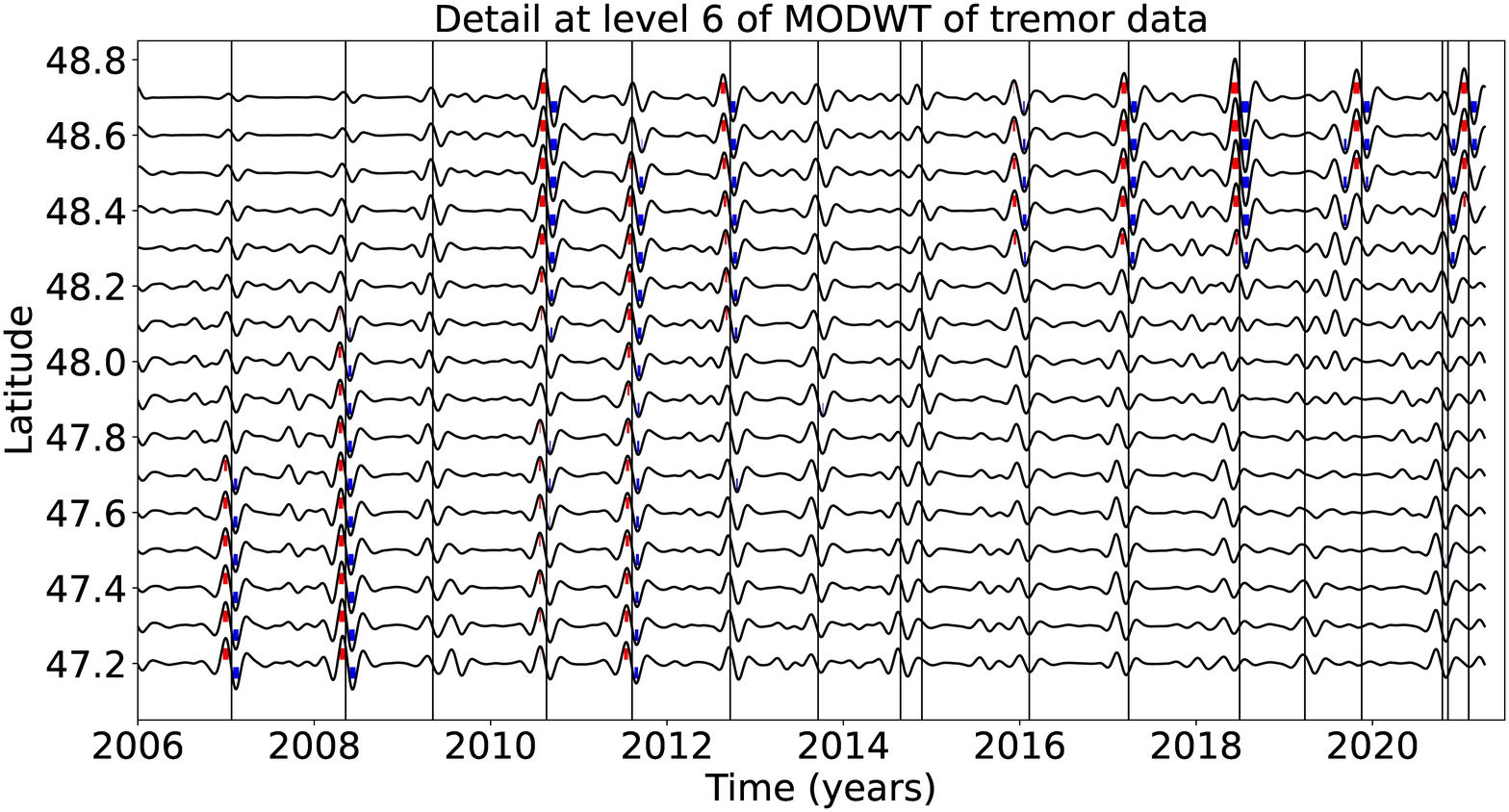}
\caption{Top: Stacked 6th level details of the wavelet decomposition of the displacement over all the GPS stations located in a 50 km radius of a given point, for the 16 red triangles indicated in Figure 3. Bottom: 6th level detail multiplied by -1 of the cumulative tremor count in a 50 km radius of a given point for the same 16 locations. The black lines represent the timings of the ETS events from Table 1. We mark by a red rectangle every time where the amplitude is higher than a threshold of 0.3 mm (for the GPS) or 0.009 (for the tremor, that is about 51 times the average value of the signal). We mark by a blue rectangle every time where the amplitude is lower than minus the threshold.}
\label{pngfiguresample}
\end{figure}

\begin{figure}
\noindent\includegraphics[width=\textwidth, trim={0cm 0cm 0cm 0cm},clip]{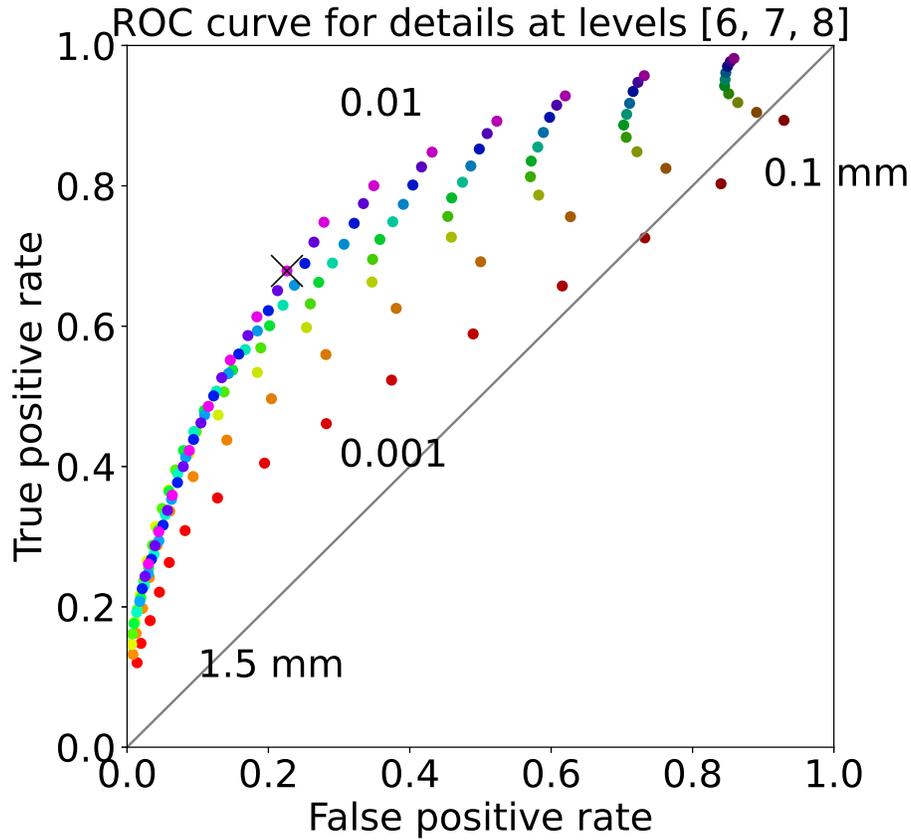}
\caption{ROC curve for the sum of the 6th, 7th, and 8th level details of the wavelet decomposition. Each dot represents the true positive rate of event detections and the false positive rate of event detections for a given pair of thresholds (for the GPS and for the tremor). The black cross marks the true positive rate and the false positive rate obtained with the thresholds used to make Figure 9. The values of the threshold are color-coded. Reds (bottom curve) correspond to the lowest value of the threshold for the tremor (0.001), while oranges, greens, blues, purples correspond to increasing values of the threshold for the tremor (up to 0.01, top curve). The brightest colors (bottom left) correspond to the highest values of the threshold for the GPS (1.5 mm), while the darker colors (top right) correspond to decreasing values of the threshold for the GPS (0.1 mm).}
\label{pngfiguresample}
\end{figure}

\begin{figure}
\noindent\includegraphics[width=\textwidth, trim={0cm 0cm 0cm 0cm},clip]{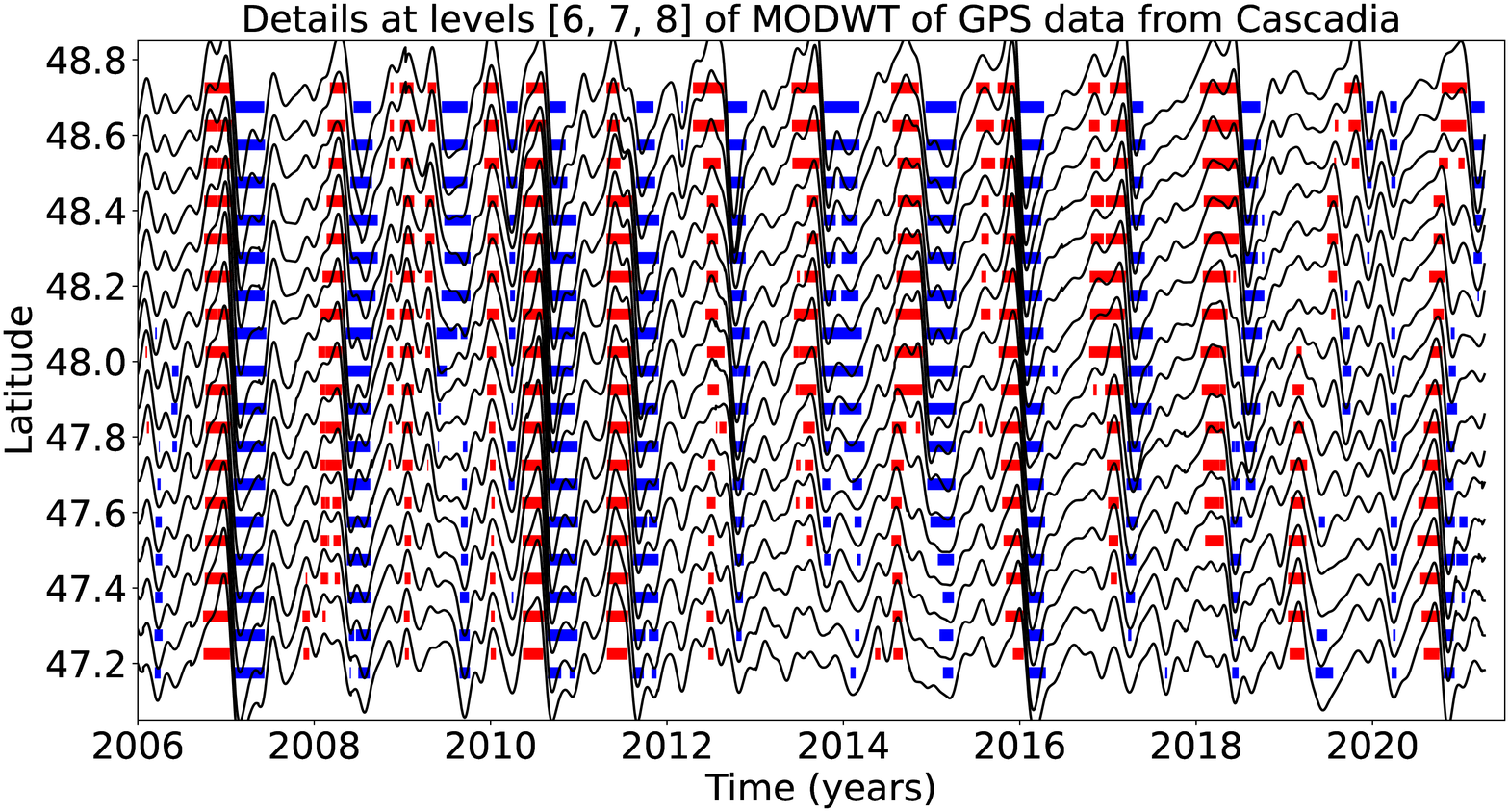}

\noindent\includegraphics[width=\textwidth, trim={0cm 0cm 0cm 0cm},clip]{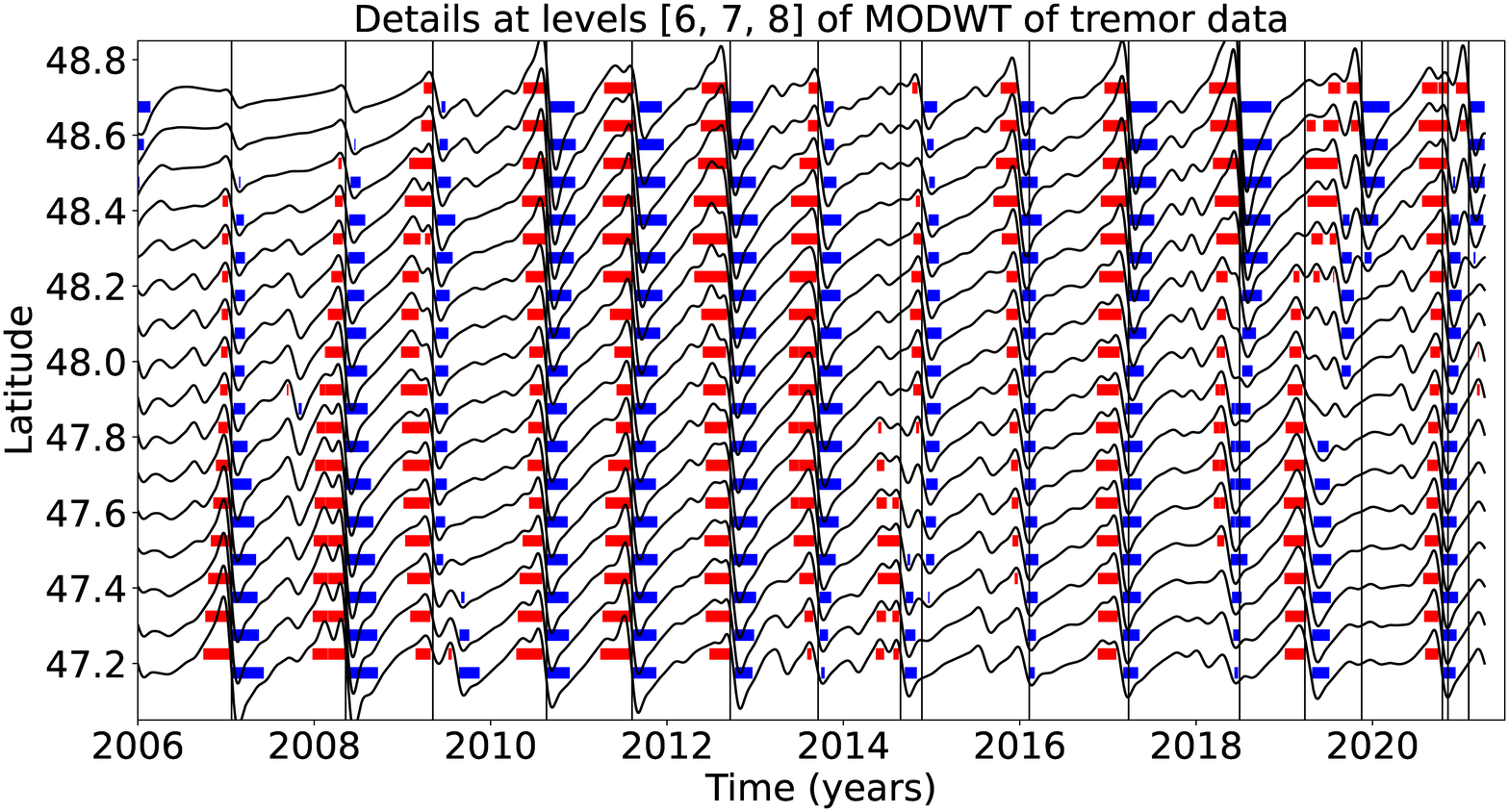}
\caption{Top: Stacked sum of the 6th, 7th and 8th levels details of the wavelet decomposition of the displacement over all the GPS stations located in a 50 km radius of a given point, for the 16 red triangles indicated in Figure 3. Bottom: Sum of the 6th, 7th and 8th levels detail multiplied by -1 of the cumulative tremor count in a 50 km radius of a given point for the same 16 locations. The black lines represent the timings of the ETS events from Table 1. We mark by a red rectangle every time where the amplitude is higher than a threshold of 0.8 mm (for the GPS) or 0.01 (for the tremor, that is about 56 times the average value of the signal). We mark by a blue rectangle every time where the amplitude is lower than minus the threshold.}
\label{pngfiguresample}
\end{figure}

\begin{figure}
\noindent\includegraphics[width=\textwidth, trim={0cm 0cm 0cm 0cm},clip]{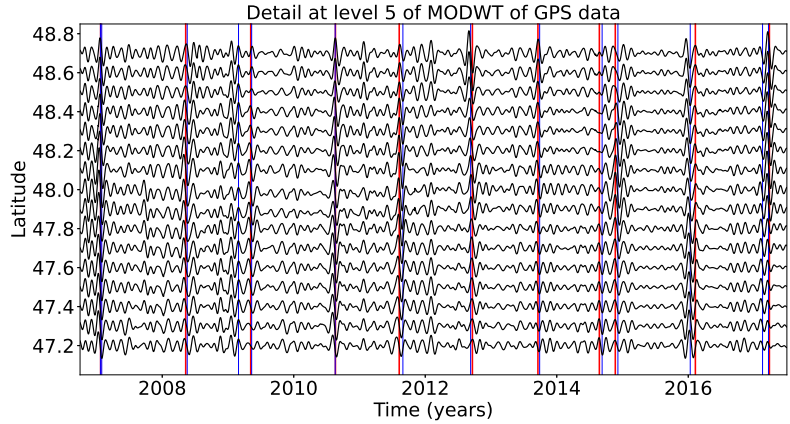}
\caption{Top: Stacked 5th level details of the wavelet decomposition of the displacement over all the GPS stations located in a 50 km radius of a given point, for the 16 red triangles indicated in Figure 3. The red lines represent the timings of the ETS events from Table 1. The blue lines represent the timings of the magnitude 5 events from the catalog of ~\citet{MIC_2019}.}
\label{pngfiguresample}
\end{figure}

\begin{figure}
\noindent\includegraphics[width=\textwidth, trim={0cm 0cm 0cm 0cm},clip]{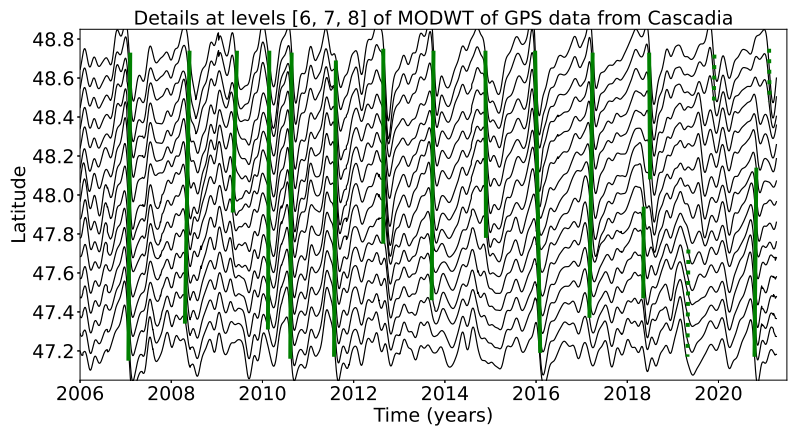}
\caption{Same as top panel of Figure 9: Stacked sum of the 6th, 7th and 8th levels details of the wavelet decomposition of the displacement over all the GPS stations located in a 50 km radius of a given point, for the 16 red triangles indicated in Figure 3. We mark with a green bar the slow slip events from Table 4 detected with the wavelet method. Full lines correspond to robust detections (1 in Table 4) and dotted lines to less robust detections (2 in Table 4).}
\label{pngfiguresample}
\end{figure}

\begin{figure}
\noindent\includegraphics[width=\textwidth, trim={0cm 0cm 0cm 0cm},clip]{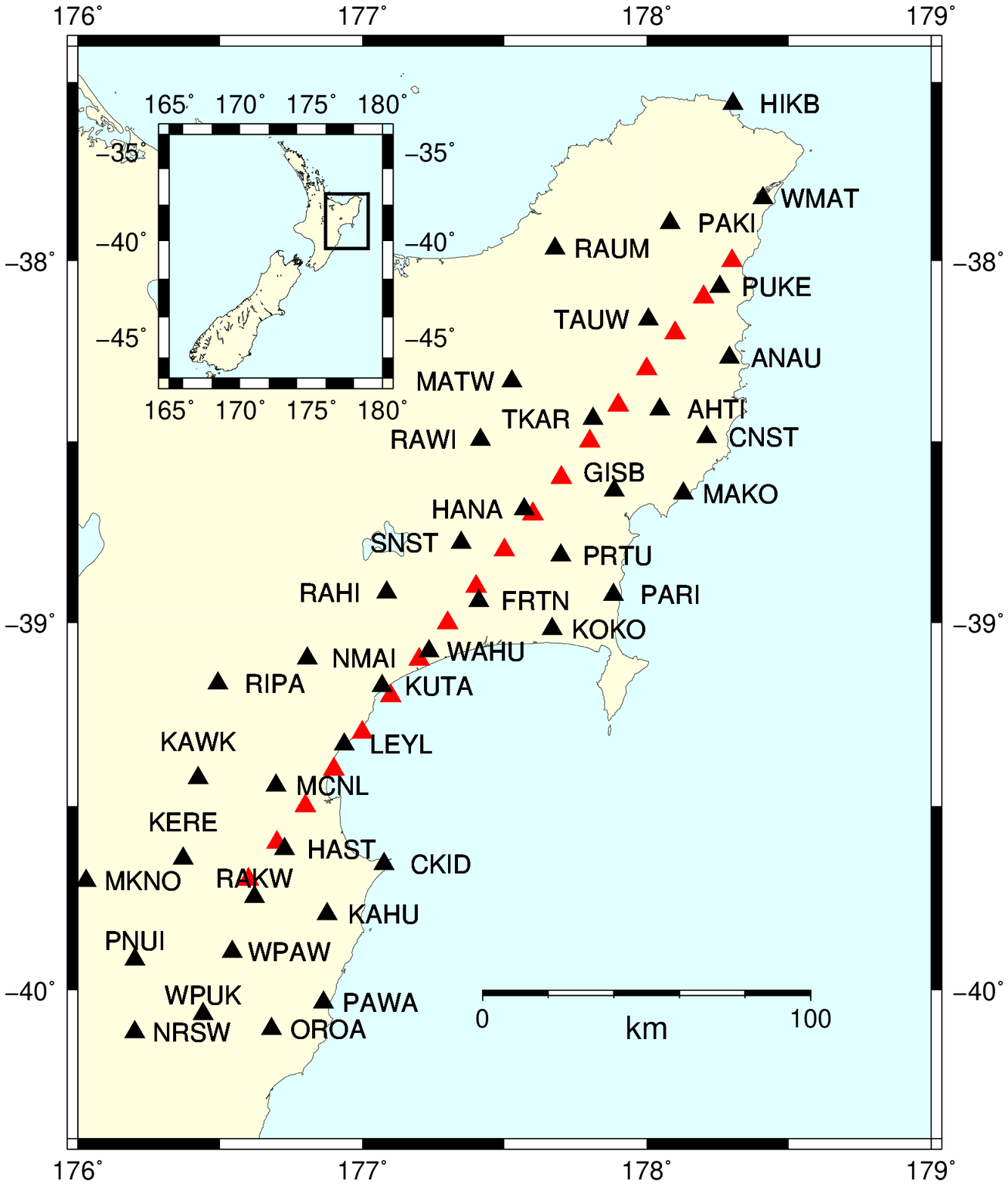}
\caption{GPS stations used for the slow slip detection in New Zealand (black triangles). The red triangles are the locations where we stack the GPS data. They are located close to the 20 km depth contour of the plate boundary from ~\citet{WIL_2013}.}
\label{pngfiguresample}
\end{figure}

\begin{figure}
\noindent\includegraphics[width=\textwidth, trim={0cm 0cm 0cm 0cm},clip]{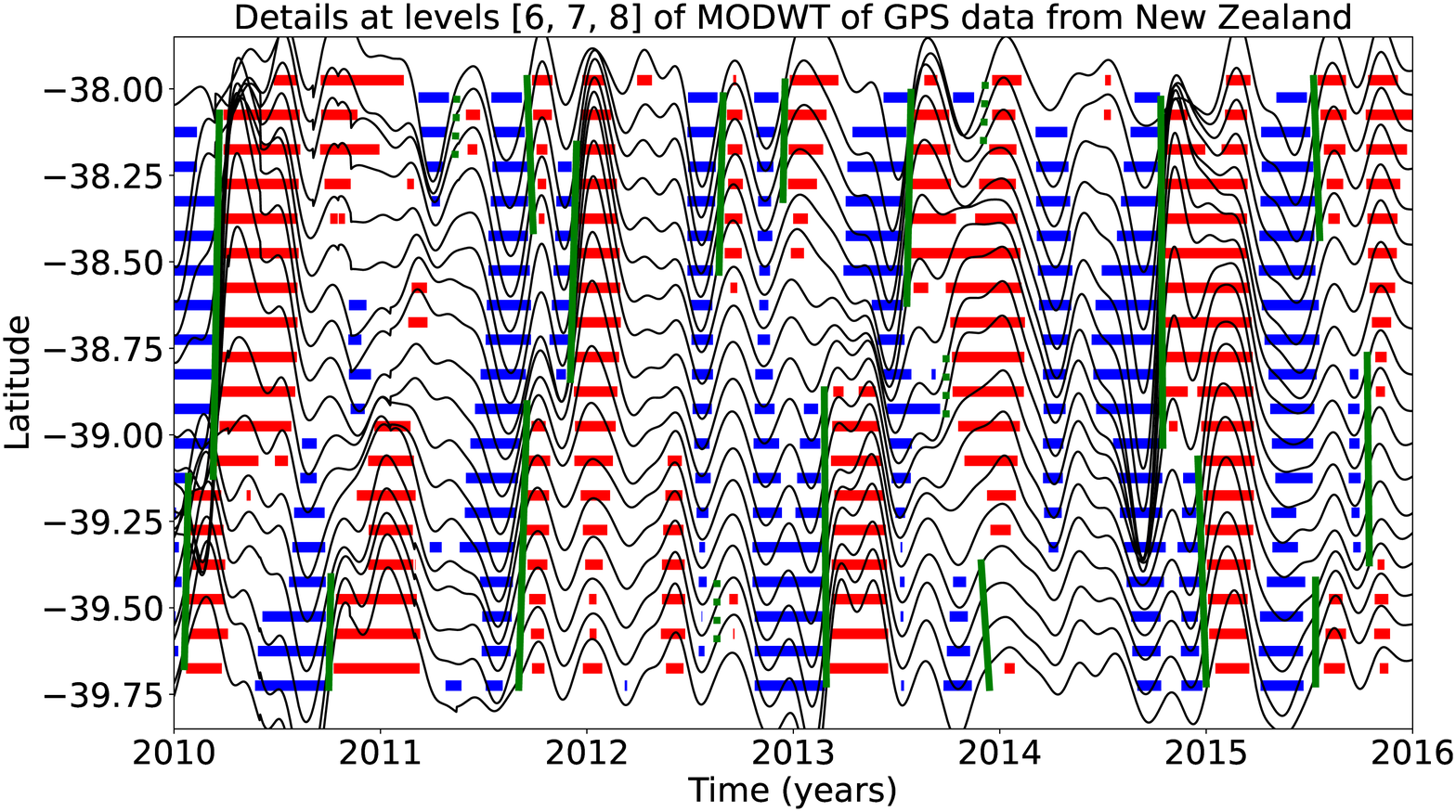}

\noindent\includegraphics[width=\textwidth, trim={0cm 0cm 0cm 0cm},clip]{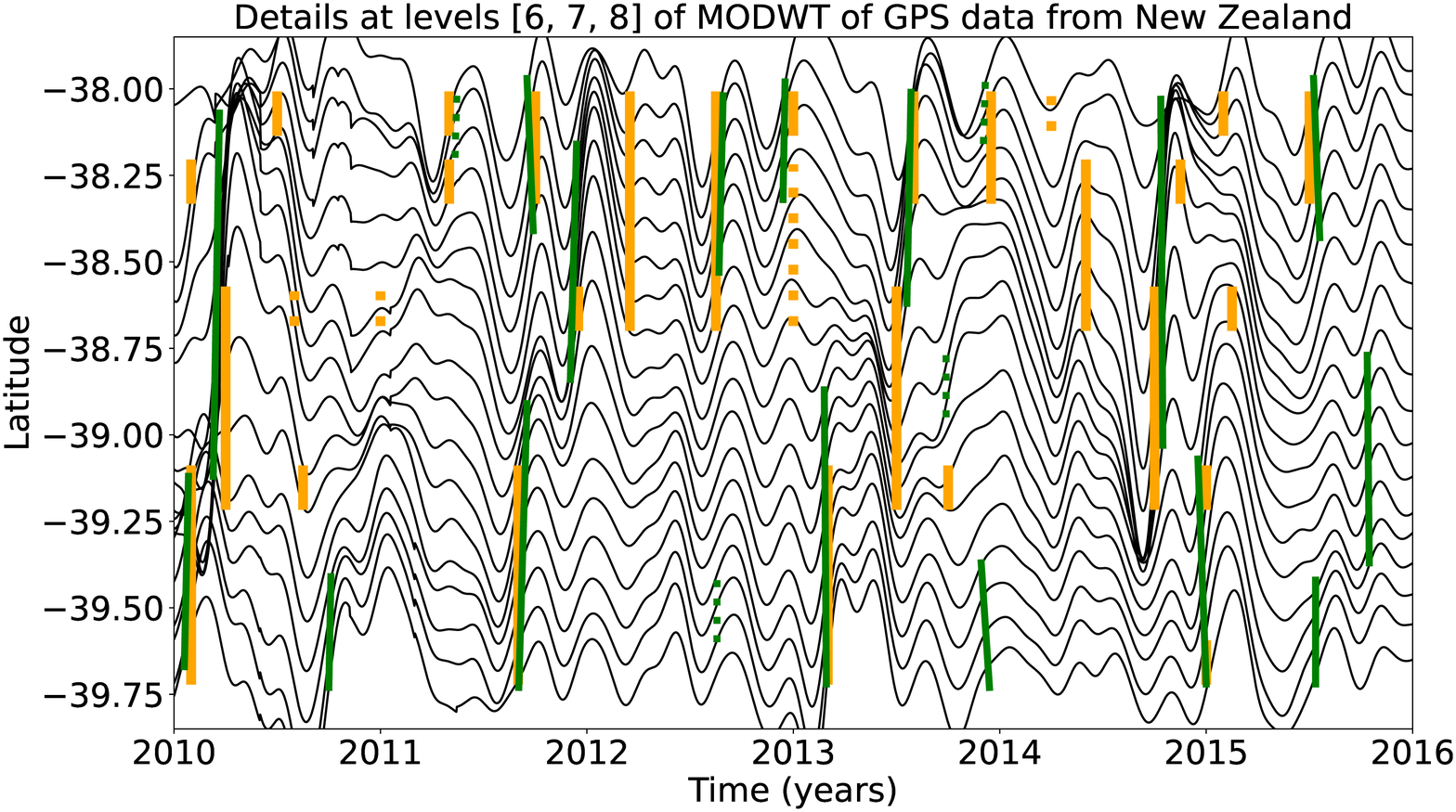}
\caption{Top: Sum of the stacked 6th, 7th and 8th level details of the wavelet decomposition of the displacement over all the GPS stations located in a 50 km radius of a given point, for the 18 red triangles indicated in Figure 12. The time period covered is 2010-2016. We mark by a red rectangle every time where the amplitude is higher than a threshold equal to 0.8 mm. We mark by a blue rectangle every time where the amplitude is lower than minus the threshold. Bottom: Sum of the stacked 6th, 7th and 8th level details of the wavelet decomposition. We mark with an orange bar the slow slip events detected by ~\citet{TOD_2016} and with a green bar the slow slip events from Table 5 detected with the wavelet method. Full lines correspond to robust detections (1 in Table 5) and dotted lines to less robust detections (2 in Table 5).}
\label{pngfiguresample}
\end{figure}

\begin{figure}
\noindent\includegraphics[width=\textwidth, trim={0cm 0cm 0cm 0cm},clip]{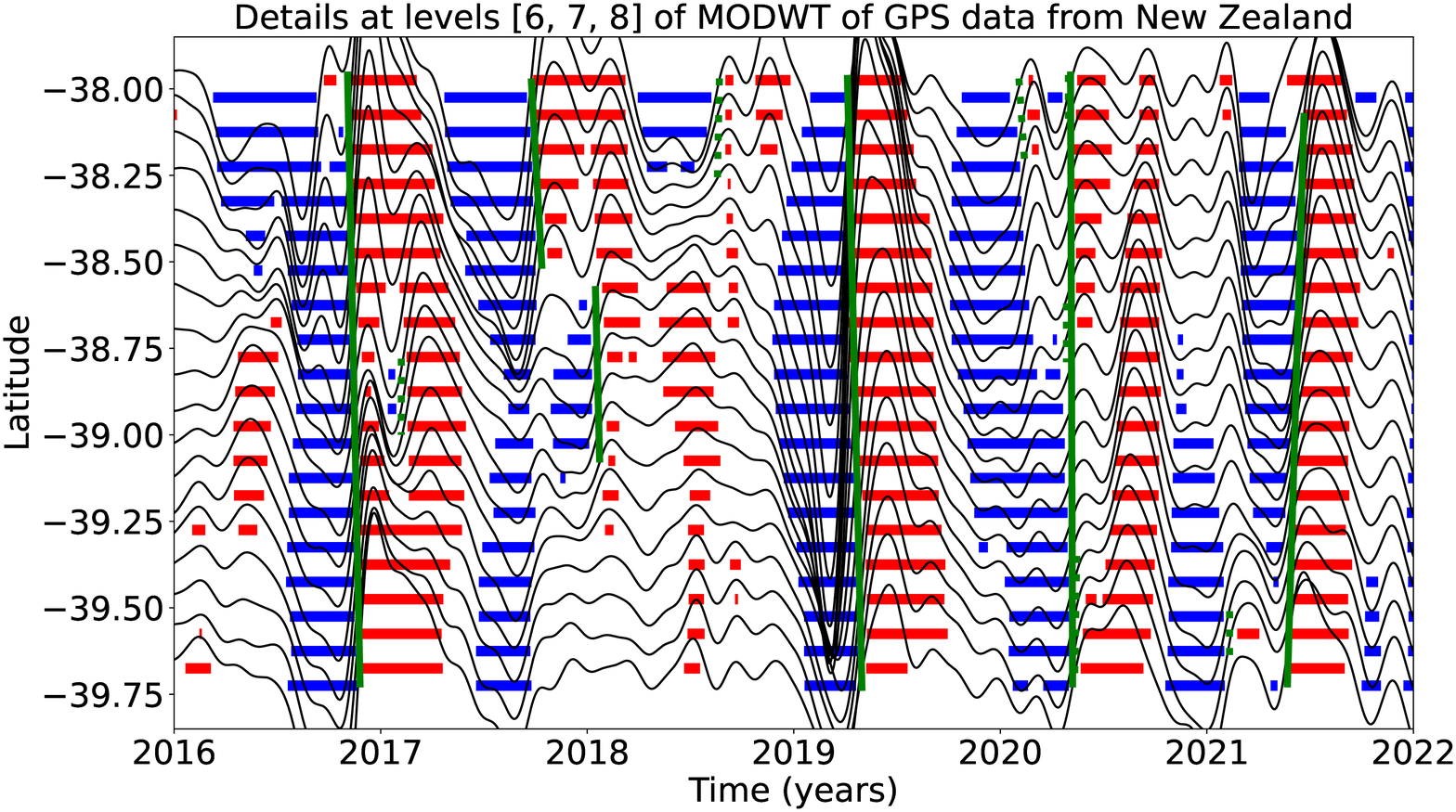}
\caption{Top: Sum of the stacked 6th, 7th and 8th level details of the wavelet decomposition of the displacement over all the GPS stations located in a 50 km radius of a given point, for the 18 red triangles indicated in Figure 12. The time period covered in 2016-2022. We mark by a red rectangle every time where the amplitude is higher than a threshold equal to 0.8 mm. We mark by a blue rectangle every time where the amplitude is lower than minus the threshold. We mark with a green bar the slow slip events from Table 6 detected with the wavelet method. Full lines correspond to robust detections (1 in Table 6) and dotted lines to less robust detections (2 in Table 6).}
\label{pngfiguresample}
\end{figure}

\end{document}